\documentclass[11pt]{article} 
\usepackage{\jobname,epsfig,amsmath,amssymb,graphics,color} 
 
\newtheorem{theorem}{Theorem}[section] 
\newtheorem{proposition}[theorem]{Proposition}
 
\newtheorem{lemma}[theorem]{Lemma}

\newcommand{\sezione}[2]{ 
\refstepcounter{section}\label{#2} 
\setcounter{equation}{0} 
\setcounter{subsection}{0} 
\addcontentsline{toc}{section} 
      {\normalsize\textbf{\thesection.\ #1}} 
\bigskip\bigskip\noindent 
\normalsize\textbf{\thesection.\ #1}\smallskip\nopagebreak} 
\def\thesection{{\arabic{section}}} 
\newcommand{\subsec}[2]{ 
\refstepcounter{subsection}\label{#2} 
\addcontentsline{toc}{subsection} 
      {\normalsize\normalfont\textit{\thesubsection.\ #1}} 
\medskip\medskip\noindent 
\normalsize\normalfont\textit{\thesubsection.\ #1}\smallskip\nopagebreak} 
\def\thesubsection{{\normalfont 
            \textit{\arabic{section}.\arabic{subsection}}}} 
\newcounter{appendice}

\newcommand{\qed}{{\hfill $\Box$}} 

\newcommand{\dis}{{\mathrm{d}}} 
\newcommand{\core}[1]{\overline{#1}} 
 
\newcommand{\const}{{\mathrm{const}}}

\def\rect#1#2{{\vcenter{\vbox{\hrule height.3pt 
            \hbox{\vrule width.3pt height#2truecm \kern#1truecm 
            \vrule width.3pt} 
            \hrule height.3pt}}}}

\newcommand{\bb}[1]{{\mathbb{#1}}} 
\newcommand{\cc}[1]{{\mathcal{#1}}} 
\newcommand{\puno}{+{\underline{1}}} 
\newcommand{\muno}{-{\underline{1}}} 
\newcommand{\eC}{C^e} 
\newcommand{\oC}{C^o} 
\newcommand{\lcrit}{\lambda} 
\newcommand{\rmT}{\mathrm{T}} 
 
\definecolor{light}{gray}{.8}

\begin{document} 
\title{Metastability for a stochastic dynamics with
a parallel heat bath updating rule} 
\author{ 
\\
{\normalsize{Emilio N.M. Cirillo$^1$ and Francesca R. Nardi$^2$}}\\ 
\\
}
\date{
\parbox{14cm}{
\renewcommand{\baselinestretch}{0.8}
\small{$^1$Dipartimento Me. Mo. Mat.,  
Universit\`a degli Studi di Roma ``La Sapienza",\\ 
$\phantom{^1}$via Antonio Scarpa 16, I--00161 Roma, Italy.\\
$\phantom{^1}$E\_mail: cirillo@dmmm.uniroma1.it
\\ \\
$^2$Eurandom, PO BOX 513, 5600MB, Eindhoven, Nl.\\
$\phantom{^2}$E\_mail: nardi@eurandom.tue.nl
}}
} 
\maketitle 
\vskip 2 cm 
\renewcommand{\baselinestretch}{1.5}
\begin{abstract} 
We consider the problem of metastability for a stochastic dynamics with a 
parallel updating rule with
single spin rates equal to those of the heat bath for the Ising 
nearest neighbors interaction.
We study the exit from the metastable phase, we describe the typical 
exit path and evaluate the exit time. 
We prove that the phenomenology 
of metastability is different from the one observed in the case of the
serial implementation of the heat bath dynamics. In particular we prove that 
an intermediate chessboard phase appears during the excursion from the minus 
metastable phase toward the plus stable phase. 
\end{abstract} 
 
\renewcommand{\baselinestretch}{1} 

 
 
 
\newpage 
\sezione{Introduction}{s:int} 
\par\noindent 
Metastable states arise when a physical system is close to a 
first order phase transition. 
If the system is prepared in the metastable phase, it takes an 
extremely long time to reach equilibrium. In physical experiments 
it is seen that if the system is not suitably perturbed 
it remains forever in the metastable phase \cite{[PL]}. 
 
A rigorous description of metastability cannot be formulated in 
terms of the standard equilibrium statistical mechanics: dynamical 
models must be considered \cite{[CGOV]}. 
The case of the stochastic serial dynamics
has been discussed, for instance, in \cite{[KO],[NS1],[S]}: 
at each step of time one of the spins on 
the lattice is updated with rates satisfying the detailed balance 
condition. In this set--up it has been seen that starting from the 
wrong metastable phase, the time needed by the system to exit the metastable 
state, namely the {\textit{exit time}}, is exponentially long in the 
inverse of the temperature. Moreover, the exit time is exactly the time 
needed to see a sufficiently large droplet, namely the 
{\textit{protocritical droplet}},  of the stable phase in the 
metastable background. Hence, the equilibrium is achieved via 
the {\textit{nucleation}} of such a protocritical seed. 
 
It is rather natural to ask oneself in which sense these results 
depend on the dynamics. 
In this paper we consider a dynamics in which  
simultaneous spin flips \cite{[BCLS],[C]} are allowed: the single 
spin flipping rates are those corresponding to the two dimensional 
nearest neighbors Ising interaction.
More precisely we study the metastable behavior of a Probabilistic Cellular
Automaton \cite{[R],[St]} which is 
reversible with respect to a Gibbs measure derived by an Hamiltonian 
with four body interaction. 
We show that the exit path from the metastable phase to the equilibrium 
changes dramatically, with respect to the serial implementation 
of the heat bath dynamics, in particular 
the system visits an intermediate metastable phase before 
reaching the equilibrium. 
This is not surprising, indeed, as it will be pointed out throughout the 
paper, there exist many deep differences between the evolution of the system 
under a serial and a parallel dynamics. 

We focus, now, on what we consider the most relevant novelty appearing in the 
study of metastability for parallel dynamics: 
in Glauber dynamics the system can jump between 
configurations differing at most for one spin, such pairs of configurations 
are called neighboring configurations. A connected domain is 
a subset of the configuration space such that for any pair of states 
it is possible to find a sequence of pairwise neighboring configurations of 
the domain joining the two states; the system, during its evolution, 
can visit the whole connected domain without exiting from the domain itself. 
In order to exit a connected domain, the system must necessarily 
cross its external boundary, that is the set of configurations not belonging 
to the domain, but having a nearest neighbor inside it. 
This sort of ``continuity" property is the key property in estimating 
the exit time, that is in establishing the minmax between the 
metastable and the stable states, 
namely the minimal energy barrier bypassed by any path 
joining the metastable to the stable state. 
 
Continuity is absent in the case of PCA's: 
any configuration is connected to any other, a path joining the metastable 
to the stable state is an arbitrary sequence of configurations starting 
with minus one and ending with plus one. The lack of continuity forces us 
to develop techniques to estimate the energy cost of any direct jump from a 
subcritical to a supercritical configuration. 
 
The paper is organized as follows: in Section \ref{s:mod} 
we define the model. In Section \ref{s:ris} we 
state our results: we first 
characterize the stable configurations (fixed points for the 
zero temperature dynamics, that is the typical droplets of the right phase 
plunged into the sea of the wrong phase); then we study the tendency 
to grow or to shrink of such droplets; finally, 
we construct the subset of the configuration space 
visited by the system in the 
metastable phase (description of the fluctuation around the 
metastable state) and, via a detailed description of the 
escape path, we estimate the exit time. 
In Sections \ref{s:dim} 
and \ref{s:dimp} we, finally, prove the Theorems and 
the Propositions. 
\par\noindent 
 
\sezione{Definition of the model}{s:mod} 
\par\noindent 
In this section we define our model, namely a  
Probabilistic Cellular Automaton reversible with respect to a four 
body hamiltonian. 
 
\subsec{Preliminary definitions}{su:pre} 
\par\noindent 
Let $\Lambda$ be a two--dimensional torus containing $L^2$ lattice sites, 
i.e., $\Lambda\subset\bb{Z}^2$ is a square containing $L^2$ points and 
having periodic boundary conditions. 
Let $\dis:(x,y)\in\Lambda\times\Lambda\rightarrow \dis(x,y)\in[0,+\infty)$ 
be the euclidean distance on the lattice $\Lambda$. 
For any $X,Y\subset\Lambda$ we define 
$\dis(X,Y):=\inf_{x\in X,y\in Y}\dis(x,y)$. 

We say that $x,y\in\Lambda$ 
are {\it nearest neighbors} iff $\dis(x,y)=1$. 
We say that the set $X\subset\Lambda$ is a {\it cluster} iff 
for any $x,y\in X$ there exist $x_1,\dots,x_k\in X$ such that 
$x_1=x$, $x_k=y$ and for any $i=1,\dots,k-1$ the two sites $x_i$ and 
$x_{i+1}$ are nearest neighbors.

Given two integer numbers $m\ge\ell\ge 1$ and $x\in\Lambda$ 
we denote by $R_{x,\ell,m}$ a 
rectangle on the dual lattice $\Lambda+(1/2,1/2)$ 
with side lengths $\ell$ and $m$ and such that $x$ is the 
first site of $\Lambda$ inside the rectangle in lexicographic order. 
We denote by 
$\core{R}_{\ell,m}:=\{x\in\Lambda:\; x\;{\mathrm{is\; inside}}\; R_{\ell,m}\}$ 
the interior of $R_{\ell,m}$. 
We will drop $x$ from the notation 
when it will be not necessary to specify the location of the rectangle on 
the lattice. We say that two rectangles $R_{x,\ell,m}$ and 
$R_{x',\ell',m'}$ 
are interacting (resp. non--interacting) iff 
$\dis(\core{R}_{x,\ell,m},\core{R}_{x',\ell',m'})\le 2$ 
(resp. $\ge\sqrt{5}$). 
 
We associate a spin variable $\sigma(x)=\pm 1$ 
to each site $x\in\Lambda$; 
the space $\{1,-1\}^\Lambda$ of 
configurations is denoted by $\cc{S}$. 
If $\sigma\in\cc{S}$ and $X\subset\Lambda$ we denote by 
$\sigma_X$ a configuration 
such that $\sigma_X(x)=\sigma(x)$ for any $x\in X$ and $\sigma_X(x)$ is 
arbitrary for any $x\in\Lambda\setminus X$. 
 
\subsec{Definition of the dynamics}{su:dinamica} 
\par\noindent 
Let $\sigma(x)=\pm 1$, for any $x\in\Lambda$, be a spin variable and 
let
\begin{equation} 
H^{(I),h}_{\Lambda}(\sigma)\equiv H^{(I)}(\sigma):= 
-\sum_{\langle x,y\rangle}\sigma(x)\sigma(y) 
-h\sum_{x\in\Lambda}\sigma(x) 
\label{hi} 
\end{equation} 
be the Ising nearest neighbors interaction, with 
the first sum performed over all the nearest neighbor pairs, 
$\sigma\in\cc{S}$ and $h\in\bb{R}$. 

Let us introduce the discrete time variable 
$n=0,1,\dots$ and denote by $\sigma_n$ the system configuration at time $n$. 
All the spins are updated simultaneously and independently at every unit 
time; the conditional probability that the spin at site $x$ takes value 
$a\in\{-1,+1\}$ at time $n$, given the configuration at time $n-1$, is 
\begin{equation} 
\begin{array}{rl} 
{\displaystyle{ 
p_x(a|\sigma_{n-1}) 
:= 
}} 
& 
{\displaystyle{ 
\frac{\exp\left\{-\beta H^{(I)} 
          (a,(\sigma_{n-1})_{\Lambda\setminus\{x\}})\right\} 
     } 
     {\exp\left\{-\beta H^{(I)} 
            (a,(\sigma_{n-1})_{\Lambda\setminus\{x\}})\right\} 
     +\exp\left\{-\beta H^{(I)} 
            (-a,(\sigma_{n-1})_{\Lambda\setminus\{x\}})\right\} 
     } 
}} 
\\ 
&\\ 
{\displaystyle{ 
= 
}} 
& 
{\displaystyle{ 
\frac{1} 
{1+\exp\left\{-2\beta a(S_{\sigma_{n-1}}(x)+h)\right\}} 
= 
\frac{1}{2} 
\left[1+a\tanh\beta \left( 
S_{\sigma_{n-1}}(x)+h\right)\right] 
}} 
\\ 
\end{array} 
\label{rule} 
\end{equation} 
where 
$\pm a,(\sigma_{n-1})_{\Lambda\setminus\{x\}}$ are the configurations 
equal to $\sigma_{n-1}$ on $\Lambda\setminus\{x\}$ and to 
$\pm a$ on $\{x\}$, 
\begin{displaymath} 
S_{\sigma}(x):= \sum_{y\in\Lambda :\;\dis(x,y)=1}\sigma(y) 
\end{displaymath} 
for any $\sigma\in\cc{S}$ and $x\in\Lambda$. 
The normalization condition 
$p_x(a|\sigma_{n-1})+p_x(-a|\sigma_{n-1})=1$ is trivially satisfied. 
Thus the time evolution is defined as 
a Markov chain on $\cc{S}$ with non--zero 
transition probabilities 
$P_\Lambda(\eta|\sigma)$ given by 
\begin{equation} 
P_\Lambda(\eta|\sigma) 
\equiv 
P_\Lambda(\sigma, \eta) 
:=\prod_{x\in\Lambda}p_x\left(\eta(x)|\sigma\right)\;\;\; 
\forall\sigma,\eta\in\cc{S}\;\;\; . 
\label{markov} 
\end{equation} 
It is straightforward \cite{[D]} 
that the above Probabilistic Cellular Automaton 
is reversible with respect to the Gibbs measure 
$
 \nu_\Lambda(\sigma):= 
   \exp\{-H_{\Lambda}(\sigma)\}/Z_{\Lambda}
$
with 
$
 Z_{\Lambda}:=\sum_{\eta\in\cc{S}}\exp\{-H_{\Lambda}(\eta)\}
$
and 
\begin{equation} 
H_{\Lambda}^{\beta,h}(\sigma)\equiv H(\sigma):= 
-\beta h\sum_{x\in\Lambda}\sigma(x) 
-       \sum_{x\in\Lambda}   \log\cosh\left[\beta 
\left( 
S_{\sigma}(x)+h\right)\right] 
\;\;\; . 
\label{ham} 
\end{equation} 
In other words the detailed balance condition 
\begin{equation} 
P_{\Lambda}(\sigma,\eta)\exp\{-H_{\Lambda}(\sigma)\}= 
P_{\Lambda}(\eta,\sigma)\exp\{-H_{\Lambda}(\eta)\} 
\label{dett} 
\end{equation} 
is satisfied for any $\sigma,\eta\in\cc{S}$. 
The interaction is short range and it is possible to extract the 
potentials: for any $\sigma\in\cc{S}$ we can write 
\begin{equation} 
\begin{array}{ll} 
{\displaystyle{ 
H(\sigma)-{\const}=}} 
& 
{\displaystyle{ 
-J_{.} 
\sum_{x\in\Lambda}\sigma(x) 
-J_{_{\langle\langle\rangle\rangle}} 
\sum_{\langle\langle x y\rangle\rangle}\sigma(x)\sigma(y) 
-J_{_{\langle\langle\langle \rangle\rangle\rangle}} 
\sum_{\langle\langle\langle x y\rangle\rangle\rangle}\sigma(x)\sigma(y) 
}} 
\\ 
&\\ 
& 
{\displaystyle{ 
-J_{_{_{\widehat{}}}} 
\sum_{\widehat{xyz}}\sigma(x)\sigma(y)\sigma(z) 
-J_{_{\diamondsuit}} 
\sum_{\diamondsuit_{xywz}}\sigma(x)\sigma(y)\sigma(w)\sigma(z)  
}} 
\\ 
\end{array} 
\label{pot} 
\end{equation} 
where the five sums are respectively performed over all the sites in $\Lambda$, 
the pairs of next to the nearest neighbors, the pairs of sites at 
distance $2$, the three site clusters composed of two consecutive not 
parallel pairs 
of next to the nearest neighbor sites and, finally, over the four site 
diamond shaped clusters. The even coupling constants are 
\begin{equation} 
\begin{array}{rl} 
{\displaystyle{ 
J_{_{\langle\langle\rangle\rangle}}= 
2\; 
J_{_{\langle\langle\langle \rangle\rangle\rangle}}}} 
& 
{\displaystyle{ 
= 
\frac{1}{8}\log\frac{\cosh\beta(4+h)\cosh\beta(4-h)} 
                     {\cosh^2(\beta h)} 
\;\; 
\stackrel{\scriptstyle\beta\to\infty}{\sim} 
\;\; 
\beta-\frac{1}{4}\beta h 
}} 
\\ 
\\ 
{\displaystyle{ 
J_{_{\diamondsuit}}}} 
& 
{\displaystyle{ 
= 
\frac{1}{16}\log\frac{\cosh\beta(4-h)\cosh^6(\beta h)\cosh\beta(4+h)} 
                     {\cosh^4\beta(2+h)\cosh^4\beta(2-h)} 
\;\; 
\stackrel{\scriptstyle\beta\to\infty}{\sim} 
\;\; 
-\frac{1}{2}\beta+\frac{3}{8}\beta h 
}} 
\\ 
\end{array} 
\label{accp} 
\end{equation} 
while the odd ones are 
\begin{equation} 
\begin{array}{rl} 
{\displaystyle{ 
J_{.}}} 
& 
{\displaystyle{ 
= 
\beta h+\frac{1}{4}\log\frac{\cosh^2\beta(2+h)\cosh\beta(4+h)} 
                            {\cosh^2\beta(2-h)\cosh\beta(4-h)} 
\;\; 
\stackrel{\scriptstyle\beta\to\infty}{\sim} 
\;\; 
\frac{5}{2}\beta h 
}} 
\\ 
\\ 
{\displaystyle{ 
J_{_{_{\widehat{}}}}}} 
& 
{\displaystyle{ 
= 
\frac{1}{16}\log\frac{\cosh^2\beta(2-h)\cosh\beta(4+h)} 
                     {\cosh^2\beta(2+h)\cosh\beta(4-h)} 
\;\; 
\stackrel{\scriptstyle\beta\to\infty}{\sim} 
\;\; 
-\frac{1}{8}\beta h 
}} 
\\ 
\end{array} 
\label{accd} 
\end{equation} 
 
\subsec{The energy and the zero temperature phase diagram}{su:fond} 
\par\noindent 
The definition of ground states is not completely trivial in our model, 
indeed the hamiltonian $H_{\Lambda}$ depends on $\beta$. 
The ground states are those configurations on which the Gibbs 
measure $\nu_{\Lambda}$ is concentrated when the limit 
$\beta\to\infty$ is considered, so they can be defined as the 
minima of the energy 
\begin{equation} 
E_{\Lambda}^{h}(\sigma)
\equiv E(\sigma):= 
\lim_{\beta\to\infty}\frac{H_{\Lambda}(\sigma)}{\beta} 
= 
-h\sum_{x\in\Lambda}\sigma(x) 
-\sum_{x\in\Lambda}|S_{\sigma}(x)+h| 
\label{hl} 
\end{equation} 
uniformly in $\sigma\in\cc{S}$. 
Notice that it is possible to write $H_{\Lambda}(\sigma)= 
\beta E_{\Lambda}(\sigma)+o(\exp\{-\beta c\})$ 
for some positive constant $c$ depending on $\sigma$. 

We consider, now, the case $h=0$: 
$E_{\Lambda}(\sigma)=-\sum_{x\in\Lambda} |S_{\sigma}(x)|$. 
It is rather clear that there exist four coexisting minima 
$\puno,\muno,\eC,\oC\in\cc{S}$: 
\begin{equation} 
\puno(x)=+1 
,\;\;\; 
\muno(x)=-1 
,\;\;\; 
\eC(x)=(-1)^{x_1+x_2} 
\;\;\;{\mathrm{and}}\;\;\;\;\; 
\oC(x)=(-1)^{x_1+x_2+1} 
\label{fond} 
\end{equation} 
for all $x=(x_1,x_2)\in\Lambda$. Notice that 
$\eC$ and $\oC$ are the chessboard configurations with plus spins 
respectively on the even and odd sublattices. 
We define $\cc{C}:=\{\oC,\eC\}$. 
 
Now, we wonder what happens when $h\not=0$: 
a full description of the zero--temperature phase diagram 
requires the introduction of a staggered magnetic field. We 
consider the new zero--temperature energy 
\begin{equation} 
E_{\Lambda}^{h_o,h_e}(\sigma):= 
-\sum_{x\in\Lambda}h_x\sigma(x) 
-\sum_{x\in\Lambda}|S_{\sigma}(x)+h_x| 
\;\;\; , 
\label{hln} 
\end{equation} 
where $h_o,h_e\in\bb{R}$ and $h_x=h_o$ (resp. $h_x=h_e$) if $x$ belongs to the 
odd (resp. to the even) sublattice. A simple calculation gives 
the energy of the four zero--field ground states: 
\begin{equation} 
\begin{array}{ll} 
{\displaystyle{ 
E_{\Lambda}^{h_o,h_e}(\puno) 
= 
-\frac{|\Lambda|}{2}\left[h_o+h_e+|4+h_o|+|4+h_e|\right]
}} 
& 
\textrm{ and }  
\;\;{\displaystyle{ 
E_{\Lambda}^{h_o,h_e}(\muno) 
= 
E_{\Lambda}^{-h_o,-h_e}(\puno) 
}}\\ 
&\\ 
{\displaystyle{ 
E_{\Lambda}^{h_o,h_e}(\oC) 
= 
-\frac{|\Lambda|}{2}\left[h_o-h_e+|4-h_o|+|4+h_e|\right]
}} 
& 
\textrm{ and }  
\;\;{\displaystyle{ 
E_{\Lambda}^{h_o,h_e}(\eC) 
= 
E_{\Lambda}^{-h_o,-h_e}(\oC) 
}}\\ 
\end{array} 
\;\;\; . 
\label{fonds} 
\end{equation} 
By comparing the four expressions (\ref{fonds}) one obtains 
the zero--temperature phase diagram in Fig. \ref{f:dia}. 
We note that on the line $h_o=h_e\equiv h$, 
depending on the sign of the magnetic field the ground state is either 
$\puno$ or $\muno$; but at $h=0$ there are four different coexisting 
ground states. 
\setlength{\unitlength}{1.3pt} 
\begin{figure} 
\begin{picture}(200,200)(-80,-50) 
\thinlines 
\put(100,50){\vector(0,1){100}} 
\put(50,100){\vector(1,0){100}} 
\thicklines 
\put(80,120){\line(1,-1){40}} 
\qbezier[10](50,120)(65,120)(80,120) 
\qbezier[10](80,150)(80,135)(80,120) 
\qbezier[10](120,80)(135,80)(150,80) 
\qbezier[10](120,80)(120,65)(120,50) 
\put(150,93){$h_o$} 
\put(103,150){$h_e$} 
\put(120,120){$\puno$} 
\put(60,130){$\eC$} 
\put(60,70){$\muno$} 
\put(130,60){$\oC$} 
\thinlines 
\end{picture} 
\vskip -4 cm 
\caption[The zero--temperature phase diagram]{The zero temperature 
phase diagram in the plane $h_o$--$h_e$. 
The four states $\puno$, $\muno$, $\oC$ and $\eC$ coexist on 
the solid line whose ending points are $(-4,4)$ and $(4,-4)$. 
Each dotted line is the boundary between two regions with 
different ground states coexisting on the line itself. 
} 
\label{f:dia} 
\end{figure}
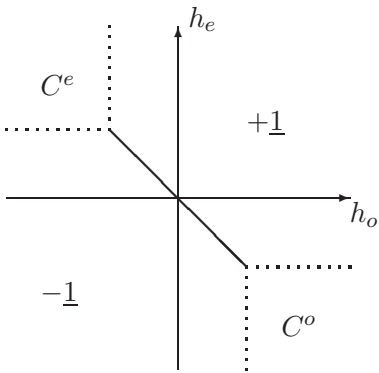 
 
\subsec{Heuristic description of the low temperature phase 
diagram}{su:tpicc} 
\par\noindent 
In this section we give a heuristic argument showing that at 
finite, but very low, temperature 
the structure of the phase diagram is not changed. More 
precisely the argument suggests that at $h=0$ the four states 
$\puno$, $\muno$, $\oC$ and $\eC$ still coexist \cite{[KV],[V]}. 

At finite temperature ground states are perturbed because small
droplets of different phases show up. The idea is to calculate the
energetic cost of a perturbation of one of the four
coexisting states via the formation of a square droplet of a
different phase. If it results that one of the four ground states
is more easily perturbed, then we will
conclude that this is the equilibrium phase at finite temperature.

A simple calculation shows that the energy cost
of a square droplet of side length $n$ of one of the two
homogeneous ground states plunged in one of the two chessboards
(or vice versa) is equal to $8n$. On the other hand if an
homogeneous phase is perturbed as above by the other homogeneous
phases, or one of the two chessboards is perturbed by the other
one, then the energy cost is $16n$.

Hence, from the energetical point of view the most convenient
excitations are those in which a homogeneous phase is perturbed
by a chessboard or vice versa. Moreover, for each state
$\muno,\puno,\eC, \oC$ there exist two possible energetically
convenient excitations: there is no entropic reason to prefer one
of the four ground states to the others when a finite low
temperature is considered. This remark strongly suggests that at
small finite temperature the four ground states still coexist.

\sezione{Results and heuristics}{s:ris} 
\par\noindent 
We pose, now, the question of metastability: 
let $h$ be positive and small; 
we prepare the system in the starting configuration $\sigma_0=\muno$ 
and we try to estimate the first time at which 
the system reaches $\puno$. 
 
The two chessboard phases coexist 
at $h=0$ with the minus and the plus phase: it is natural to wonder if 
these phases play a role during the escape from the minus metastable phase 
toward the plus stable phase when the external magnetic field is positive 
and small. 
 
The main feature of PCA models is that the system can jump from any 
configuration to any other, in contrast with what happens in serial 
Glauber dynamics, where transitions are allowed only between configurations 
differing at most for one spin. We remark that in this model the single 
spin flip is not a local event, in the sense that its probability depends 
on all the spin of the lattice. Indeed, 
given $X\subset\Lambda$ we denote by 
$\sigma^{X}$ the configuration obtained by 
flipping in $\sigma$ all the spins at sites $x\in X$; 
if $X=\{x\}$ for some $x\in\Lambda$, then by abuse of notation we will 
denote $\sigma^X=\sigma^{\{x\}}=\sigma^x$. 
Now, by (\ref{markov}) we have that 
\begin{equation} 
P_{\Lambda}(\sigma,\sigma^X)= 
\prod_{x\in X} p_{x}\left(\sigma^{x}(x)|\sigma\right) 
\prod_{y\in\Lambda\setminus X} p_{y}\left(\sigma(y)|\sigma\right) 
=\prod_{x\in X} p_{x}\left(-\sigma(x)|\sigma\right) 
\prod_{y\in\Lambda\setminus X} p_{y}\left(\sigma(y)|\sigma\right) 
\;\;\; , 
\label{flipm} 
\end{equation} 
that is the probability to flip the spins inside $X$ 
depends also on the probability that spins  
outside $X$ are not flipped. Notice that this is true even if $|X|=1$, 
namely if only one spin is flipped. 
 
\subsec{Stable configurations}{su:stabili} 
\par\noindent 
First of all we characterize 
the stable configurations of the system, namely those configurations 
$\sigma\in\cc{S}$ such that  
$P_{\Lambda}(\sigma,\sigma)\to 1$ in the limit $\beta\to\infty$.  
Equivalently, $\sigma\in\cc{S}$ is a stable configuration if and only if
$P_{\Lambda}(\sigma,\eta)\to 0$ in the limit $\beta\to\infty$ for all 
$\eta\in\cc{S}\setminus\{\sigma\}$.
 
We discuss, now, the possible single spin events. 
In Table \ref{tab:sin} we consider a site $x$ and we draw all the possible 
configurations in a five spin cross centered at $x$. The probability 
$p_x(+1|\sigma)$ to see $+1$ at $x$ is evaluated 
(we recall $p_x(-1|\sigma)=1-p_x(+1|\sigma)$). From Table \ref{tab:sin} 
it is clear that in the limit $\beta\to\infty$ the 
probability associated to a single spin event is either one or zero, in the 
sequel we will respectively say {\it high} and {\it low probability events}. 
By (\ref{flipm}) 
it follows that the same limiting behavior is valid 
in general for any transition $P_{\Lambda}(\sigma,\sigma^X)$ with 
$\sigma\in\cc{S}$ and $X\subset\Lambda$; in this sense PCA's are 
a generalization of deterministic Cellular Automata.
We remark that for 
any $\sigma\in\cc{S}$ 
there exists a unique configuration 
$\eta\in\cc{S}$ such that 
the transition $\sigma\to\eta$ happens with high probability, that is 
$P_{\Lambda}(\sigma,\eta) 
\; 
\stackrel{\scriptstyle\beta\to\infty}{\longrightarrow}\; 
1$. 
We note, moreover, that $\eta=\rmT\sigma$, where 
$\rmT:\sigma\in\cc{S}\longrightarrow 
\rmT\sigma\in\cc{S}$ is the map such that for each $x\in\Lambda$
\begin{equation} 
\rmT\sigma(x):=\left\{ 
\begin{array}{ll} 
\sigma^x(x)\;\;\;\;&{\mathrm{if}}\;\;p_x(\sigma^x(x)|\sigma)\; 
\stackrel{\scriptstyle\beta\to\infty}{\longrightarrow}\; 1\\ 
&\\ 
\sigma(x)\;\;\;\;&{\mathrm{if}}\;\;p_x(\sigma^x(x)|\sigma)\; 
\stackrel{\scriptstyle\beta\to\infty}{\longrightarrow}\; 0\\ 
\end{array} 
\right. 
\label{unico} 
\end{equation} 
that is at each site we do the right thing in the sense 
of following the drift. We can say that $\sigma\in\cc{S}$ 
is a stable configuration iff $\sigma=\rmT\sigma$. 
\begin{table} 
\begin{center} 
\begin{tabular}{llll} 
\begin{tabular}{ccc} 
 &$-$& \\ 
$-$&$x$&$-$\\ 
 &$-$& \\ 
\end{tabular} 
& 
$\frac{1}{1+e^{2\beta(4-h)}}\simeq e^{-2\beta(4-h)}$ 
& 
\begin{tabular}{ccc} 
 &$-$& \\ 
$-$&$x$&$+$\\ 
 &$-$& \\ 
\end{tabular} 
& 
$\frac{1}{1+e^{2\beta(2-h)}}\simeq e^{-2\beta(2-h)}$ 
\\ 
&&&\\ 
\begin{tabular}{ccc} 
 &$-$& \\ 
$-$&$x$&$+$\\ 
 &$+$& \\ 
\end{tabular} 
& 
$\frac{1}{1+e^{-2\beta h}}\simeq 1-e^{-2\beta h}$ 
& 
\begin{tabular}{ccc} 
 &$-$& \\ 
$+$&$x$&$+$\\ 
 &$+$& \\ 
\end{tabular} 
& 
$\frac{1}{1+e^{-2\beta(2+h)}}\simeq 1-e^{-2\beta(2+h)}$ 
\\ 
&&&\\ 
\begin{tabular}{ccc} 
 &$+$& \\ 
$+$&$x$&$+$\\ 
 &$+$& \\ 
\end{tabular} 
& 
$\frac{1}{1+e^{-2\beta(4+h)}}\simeq 1-e^{-2\beta(4+h)}$ 
&&\\ 
\end{tabular} 
\end{center} 
\caption[Single spin--events probabilities]{Probabilities for single spin 
events: probability to see $+1$ at site $x$ at time $t$, with the 
neighboring configuration at time $t-1$ drawn in the picture.} 
\label{tab:sin} 
\end{table} 

In order to characterize the stable states of the model we 
need few more definitions: let $C\in\cc{C}=\{\oC,\eC\}$, we 
denote by $\cc{S}_C\subset\cc{S}$ the set of configurations 
with a well defined sea of chessboard $C$. 
Similarly we define $\cc{S}_{\muno}, \cc{S}_{\puno}\subset\cc{S}$ and 
we set $\cc{S}_{\cc{C}}:=\cc{S}_{\oC}\cup\cc{S}_{\eC}$. 
More precisely, for each $\alpha\in\{\muno,\puno\}$, for each
$\sigma\in\cc{S}_{\alpha}$ there esists a percolating cluster 
$X\subset\Lambda$ such that $\sigma_X=\alpha_X$ and 
$\sigma_X=(\rmT^n\sigma)_X$ for all $n\ge 1$;
for each $\alpha\in\{\eC,\oC\}$, for each
$\sigma\in\cc{S}_{\alpha}$ there esists a percolating cluster 
$X\subset\Lambda$ such that $\sigma_X=\alpha_X$ and 
$\sigma_X=(\rmT^{2n}\sigma)_X$ for all $n\ge 1$. 

\begin{proposition} 
\label{p:std} 
A configuration $\sigma\in\cc{S}_{\muno}$ is 
stable for the PCA (\ref{markov}) iff 
$\sigma(x)=+1$ for all the 
sites $x$ inside a collection of pairwise non--interacting 
rectangles of minimal side length $\ell\ge 2$ and $\sigma(x)=-1$ elsewhere. 
A configuration $\sigma\in\cc{S}_{\puno}$ is stable iff 
$\sigma=\puno$. 
There is no stable configuration $\sigma\in\cc{S}_{\cc{C}}$. 
\end{proposition} 
\par\noindent 
In other words we can say that the only not trivial 
stable states are configurations with well separated rectangular 
droplets of pluses inside the sea of minuses. 
The Proposition \ref{p:std} follows from \cite{[NS1]} 
and Lemma \ref{l:sta}. 
\begin{lemma} 
\label{l:sta} 
A configuration $\sigma\in\cc{S}$ is 
stable for the PCA (\ref{markov}) iff 
\begin{displaymath} 
p_x(\sigma^x(x)|\sigma) 
\; 
\stackrel{\scriptstyle\beta\to\infty}{\longrightarrow}\; 
0 
\;\;\forall x\in\Lambda 
\end{displaymath} 
\end{lemma} 
\addcontentsline{toc}{subsubsection}{{\it Proof of Lemma \ref{l:sta}}} 
\noindent{\it Proof of Lemma \ref{l:sta}.}\/ 
Suppose
$p_x(\sigma^x(x)|\sigma)
\stackrel{\scriptstyle\beta\to\infty}{\longrightarrow}\;0$ for all 
$x\in\Lambda$:
let $\eta\in\cc{S}\setminus\{\sigma\}$, there exists 
$X\subset\Lambda$ and $X\not=\emptyset$ 
such that $\eta=\sigma^X$; thus, by equation 
(\ref{flipm}) one has 
\begin{displaymath} 
P_{\Lambda}(\sigma,\eta)= 
P_{\Lambda}(\sigma,\sigma^X)= 
\prod_{x\in X} p_{x}(\sigma^{x}(x)|\sigma) 
\prod_{y\in\Lambda\setminus X} p_{y}(\sigma(y)|\sigma) 
\; 
\stackrel{\scriptstyle\beta\to\infty}{\longrightarrow}\; 
0 
\end{displaymath} 
Suppose $\sigma$ is a stable configuration:
$P(\sigma,\sigma)\rightarrow 1$ in the limit $\beta\to\infty$, 
(\ref{markov}) and the normalization condition 
$p_x(\sigma(x)|\sigma)+p_x(\sigma^x(x)|\sigma)=1$ imply the statement.
\qed\smallskip 
 
\subsec{Stable pairs and traps}{su:ris} 
\par\noindent 
The configurations in which our system can be trapped are not exhausted 
by the stable configurations. Indeed, 
let $\sigma\in\cc{S}$ and $\eta=\rmT\sigma\not=\sigma$ 
the unique state reached with high probability starting from $\sigma$. 
If it were $\rmT\eta=\sigma$, then the system would jump 
back and forth from $\sigma$ to $\eta$ with probability going to one 
in the zero temperature limit; the system would be trapped into a two 
state loop. Given $\sigma,\eta\in\cc{S}$ and $\sigma\not=\eta$, 
we say that they form a 
``stable pair" iff 
$\eta=\rmT\sigma$ and $\rmT\eta=\sigma$. 
The two chessboard configurations $\oC$ and $\eC$ are a 
simple example of a stable pair. 

We discuss two important properties of the stable pairs.
{}From the detailed balance condition 
(\ref{dett}) it follows that if $\sigma,\eta\in\cc{S}$ form 
a stable pair, then they have the same energy, namely 
$E_{\Lambda}(\sigma)=E_{\Lambda}(\eta)$. Indeed, from 
(\ref{dett}) and (\ref{hl}) 
we have $E_{\Lambda}(\sigma)-E_{\Lambda}(\eta)= 
\lim_{\beta\to\infty}[H_{\Lambda}(\sigma)- 
                      H_{\Lambda}(\eta)]/\beta= 
\lim_{\beta\to\infty}(1/\beta)\log[P_{\Lambda}(\sigma,\eta)/ 
                                   P_{\Lambda}(\eta,\sigma)]$. Now, 
the fact that $\sigma$ and $\eta$ form a stable pair implies
$\lim_{\beta\to\infty}P_{\Lambda}(\sigma,\eta)= 
 \lim_{\beta\to\infty}P_{\Lambda}(\eta,\sigma)=1$; hence 
$E_{\Lambda}(\sigma)=E_{\Lambda}(\eta)$. By using results in 
Table \ref{tab:sin} one can show that $H_{\Lambda}(\sigma)$ 
and $H_{\Lambda}(\eta)$ differ for a quantity exponentially small in 
$\beta$. 

The remark above and the detailed balance condition suggests 
that the system cannot be trapped in loops longer than two. 
Indeed, consider a sequence 
$\sigma_1,\dots,\sigma_n\in\cc{S}$ such that $\sigma_{i+1}=\rmT\sigma_i$
for all $i=1,\dots,n-1$, and suppose, by absurdity, that 
$\sigma_1=\rmT\sigma_n$. The property above implies that 
either $E_{\Lambda}(\sigma_1)-E_{\Lambda}(\sigma_n)=c>0$ or 
$E_{\Lambda}(\sigma_1)-E_{\Lambda}(\sigma_n)=0$. In the first case from 
the detailed balance we get 
$
 |\log[P_{\Lambda}(\sigma_1,\sigma_n)/P_{\Lambda}(\sigma_n,\sigma_1)]
 -c\beta|\to 0
$ 
in the limit $\beta\to\infty$; hence using the hypothesis 
$\sigma_1=\rmT\sigma_n$ we easily get an absurd.
In the second case the detailed balance implies 
$
 |\log[P_{\Lambda}(\sigma_1,\sigma_n)/P_{\Lambda}(\sigma_n,\sigma_1)]
 |\rightarrow 0
$,
that, togheter with $\sigma_1=\rmT\sigma_n$, gives
$P_{\Lambda}(\sigma_1,\sigma_n)\to 1$, which is absurd because by 
hypothesis we have $\sigma_2=\rmT\sigma_1$.

We say that $\sigma\in\cc{S}$ is a {\it trap} if either $\sigma$ is 
a stable configuration or the pair $(\sigma,\rmT\sigma)$ is a 
stable pair. We also let $\cc{M}\subset\cc{S}$ the collection of 
all the traps. 
Now, we give a full description of the 
stable pairs in 
$\cc{S}_{\puno}\cup\cc{S}_{\cc{C}}\cup\cc{S}_{\muno}$ 
(see Fig. \ref{f:ris}): 
the most general stable pair living 
in a sea of minus is made of rectangular flip--flopping droplets of chessboard 
plunged in the sea of minuses and well separated stable droplets of 
pluses living inside the sea of minuses or inside a chessboard droplet. 
\begin{proposition} 
\label{p:res} 
$i)$ For any $\sigma\in\cc{S}_{\puno}\setminus\{\puno\}$ 
the pair $(\sigma,\rmT\sigma)$ is not a stable pair. 
$ii)$ Given $C\in\cc{C}$ and 
$\sigma\in\cc{S}_C$ the pair $(\sigma,\rmT\sigma)$ is a 
stable pair iff 
there exist $k\ge 0$ pairwise non--interacting rectangles 
$R_{\ell_1,m_1}$, $R_{\ell_2,m_2}$, \dots, $R_{\ell_k,m_k}$, 
such that $2\le\ell_i\le m_i\le L-2$ for any $i=1,\dots,k$, 
$\sigma_\Re=\puno_\Re$ 
($\sigma$ 
coincides with $\puno$ inside the rectangles) 
and $\sigma_{\Lambda\setminus\Re}=C_{\Lambda\setminus\Re}$ 
($\sigma$ coincides with the chessboard $C$ 
outside the rectangles), where $\Re:=\bigcup_{i=1}^k\core{R}_{\ell_i,m_i}$. 
$iii)$ 
Given $\sigma\in\cc{S}_{\muno}$ the pair $(\sigma,\rmT\sigma)$ is a 
stable pair iff 
there exist $k\ge 1$ rectangles 
$R_{\ell_1,m_1}$, $R_{\ell_2,m_2}$, \dots, $R_{\ell_k,m_k}$, 
with $2\le\ell_i\le m_i\le L-2$ for any $i=1,\dots,k$, and 
there exists an integer $s\in\{1,\dots,k\}$ such that the 
following conditions are fulfilled: 
\begin{enumerate} 
\item 
$\core{R}_{\ell_i,m_i}\cap\core{R}_{\ell_j,m_j}=\emptyset$ 
and $\ell_i\ge 2$ for any $i,j\in\{1,\dots,k\}$; 
\item 
for any $j\in\{1,\dots,s\}$ the family 
$\{R_{\ell_j,m_j},R_{\ell_{s+1},m_{s+1}},\dots,R_{\ell_k,m_k}\}$ 
is a family of pairwise non--interacting rectangles; 
\item 
$\sigma_{\Lambda\setminus\Re}=\muno_{\Lambda\setminus\Re}$ 
where $\Re:=\bigcup_{i=1}^k\core{R}_{\ell_i,m_i}$ 
($\sigma$ coincides with $\muno$ outside the rectangles); 
\item 
$\sigma_{\core{R}_{\ell_j,m_j}}= 
 \puno_{\core{R}_{\ell_j,m_j}}$ 
for any $j\in\{s+1,\dots,k\}$ 
($\sigma$ is plus inside 
$R_{\ell_{s+1},m_{s+1}}$,\dots,$R_{\ell_k,m_k}$); 
\item 
for any $j\in\{1,\dots,s\}$ 
there exist $k'\equiv k'(j)\ge 0$ rectangles 
$R'_{\ell'_1,m'_1}=R'_{\ell'_1,m'_1}(j)$, \dots, 
$R'_{\ell'_{k'},m'_{k'}}=R'_{\ell'_{k'},m'_{k'}}(j)$ 
such that the following conditions are fulfilled: 
\newcounter{reson1} 
\begin{list} 
{\arabic{enumi}.\arabic{reson1}.}{ 
\usecounter{reson1} 
\setlength{\labelwidth}{2cm} 
} 
\item 
$\core{R'}_{\ell'_i,m'_i}\subset\core{R}_{\ell_j,m_j}$ 
for any $i\in\{1,\dots,k'\}$; 
\item 
for any $j=1,\dots,s$ 
the family $\{R'_{\ell'_i,m'_i}:\; i=1,\dots,k'\}$ (recall 
$R'_{\ell'_i,m'_i}=R'_{\ell'_i,m'_i}(j)$ for any $i=1,\dots,k'=k'(j)$) 
is 
a family of pairwise non--interacting rectangles; 
\item 
$\sigma_{\Re'}=\puno_{\Re'}$ 
where $\Re'\equiv\Re'(j):=\bigcup_{i=1}^{k'}\core{R'}_{\ell'_i,m'_i}$ 
\item 
either $\sigma_{\core{R}_{\ell_j,m_j}\setminus\Re'}= 
           \oC_{\core{R}_{\ell_j,m_j}\setminus\Re'}$ 
or     $\sigma_{\core{R}_{\ell_j,m_j}\setminus\Re'}= 
           \eC_{\core{R}_{\ell_j,m_j}\setminus\Re'}$; 
\end{list} 
\item 
for any $i,j\in\{1,\dots,s\}$ the two rectangles 
$R_{\ell_j,m_j}$ and $R_{\ell_i,m_i}$ must be non--interacting if 
$\sigma_{\core{R}_{\ell_j,m_j}\setminus\Re'(j)}= 
 \sigma_{\core{R}_{\ell_i,m_i}\setminus\Re'(i)}$. 
\end{enumerate} 
\end{proposition} 
\setlength{\unitlength}{1pt} 
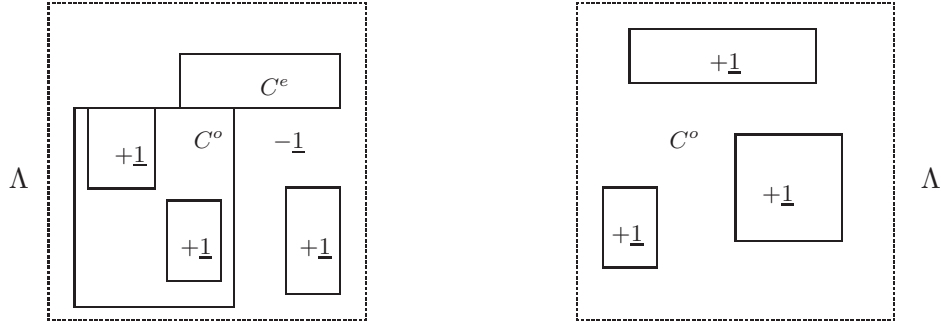
\begin{figure} 
\begin{picture}(200,200)(-190,-40) 
\thinlines 
\put(-100,50){\dashbox(120,120){}} 
\put(-115,100){$\Lambda$} 
\put(-15,115){{\footnotesize{$\muno$}}} 
\put(-20,135){{\footnotesize{$\eC$}}} 
\put(-5,75){{\footnotesize{$\puno$}}} 
\put(-45,115){{\footnotesize{$\oC$}}} 
\put(-75,110){{\footnotesize{$\puno$}}} 
\put(-50,75){{\footnotesize{$\puno$}}} 
\put(-90,55){\framebox(60,75){}} 
\put(-50,130.5){\framebox(60,20){}} 
\put(-10,60){\framebox(20,40){}} 
\put(-90,55){\framebox(60,75){}} 
\put(-85,100){\framebox(25,30){}} 
\put(-55,65){\framebox(20,30){}} 
\put(100,50){\dashbox(120,120){}} 
\put(230,100){$\Lambda$} 
\put(135,115){{\footnotesize{$\oC$}}} 
\put(113,80){{\footnotesize{$\puno$}}} 
\put(150,145){{\footnotesize{$\puno$}}} 
\put(170,95){{\footnotesize{$\puno$}}} 
\put(110,70){\framebox(20,30){}} 
\put(120,140){\framebox(70,20){}} 
\put(160,80){\framebox(40,40){}} 
\end{picture} 
\vskip -3 cm  
\caption[The most general stable pair]{On the left (resp. right) 
the most general $\sigma\in\cc{S}_{\muno}$ (resp. $\cc{S}_C$) such that 
$(\sigma,\rmT\sigma)$ is a stable pair.} 
\label{f:ris} 
\end{figure} 
 
\subsec{Basic tools}{su:strum} 
\par\noindent 
In this Section we discuss the main tools that will be used in the  
following: first of all we notice that in our model   
the difference of energy between two configurations 
$\sigma,\eta\in\cc{S}$ is not sufficient to say if the system prefers 
to jump from $\sigma$ to $\eta$ or vice versa. Indeed, 
there exist pairs of configurations 
$\sigma,\eta\in\cc{S}$ such that the system sees a sort of energetic 
barrier both in the $\sigma\to\eta$ and in the 
$\eta\to\sigma$ transition. 
Let us define a sort 
of ``communicating height" $H(\sigma,\eta)$ 
for each pair $(\sigma,\eta)\in\cc{S}\times\cc{S}$ of the configuration space 
such that 
\begin{equation} 
P_{\Lambda}(\sigma,\eta)=:e^{-[H(\sigma,\eta)-H(\sigma)]} 
\;\;\; . 
\label{nh} 
\end{equation} 
More precisely, we consider a new hamiltonian 
$H:\cc{S}\times\cc{S}\cup\cc{S}\longrightarrow \bb{R}$ 
defined as in (\ref{ham}) for any $\sigma\in\cc{S}$ and such that 
\begin{equation} 
H(\sigma,\eta):=H(\sigma)-\log P_{\Lambda}(\sigma,\eta) 
\;\;\; . 
\label{nhe} 
\end{equation} 
Note that, by virtue of the detailed balance principle 
(\ref{dett}), we have $H(\sigma,\eta)=H(\eta,\sigma)$. 
Remark: if either $P_{\Lambda}(\sigma,\eta)$ or 
$P_{\Lambda}(\eta,\sigma)$ tends to zero in the limit $\beta\to\infty$ 
then $H(\sigma,\eta)=\max\{H(\sigma),H(\eta)\}+o(\exp\{-\beta c\})$, for 
some strictly positive constant $c$; in other words in these cases 
the energetic barrier seen by the system is exactly 
the difference of energy between the two configurations. 
 
We notice that in \cite{[OS]} 
it has already been remarked that the communicating 
heights allow to define 
the most general kind of reversible dynamics (see Section 3 
in \cite{[OS]}). Now 
we want to restate in this setup some of the results of \cite{[OS]} that will 
be our basic tools in next sections. 
 
We say that a configuration $\sigma\in\cc{S}$ is a local minimum of the 
energy iff 
$H(\sigma,\eta)-H(\sigma)>0$ 
for any $\eta\in\cc{S}\setminus\{\sigma\}$. 
The local minima of the energy are nothing but the stable configurations 
defined above. 
A sequence of configurations $\omega=\{\omega_0,\dots,\omega_n\}$ is called a 
``path"; $|\omega|$ is the number of configurations in the path. 
We call ``height along the path $\omega$" the real 
number 
\begin{equation} 
\Phi_{\omega}:= 
\max_{i=1,\dots,|\omega|} 
H(\omega_{i-1},\omega_i) 
\;\;\; . 
\label{alpa} 
\end{equation} 
Given two configurations $\sigma,\eta\in\cc{S}$ 
we denote by $\Theta(\sigma,\eta)$ the set of all the paths 
$\omega=\{\omega_0,\dots,\omega_n\}$ such that 
$\omega_0=\sigma$ and $\omega_n=\eta$. 
The ``minimal height" (minmax) between $\sigma$ and $\eta$ 
is defined as 
\begin{equation} 
\Phi(\sigma,\eta):= 
\min_{\omega\in\Theta(\sigma,\eta)}\;\Phi_{\omega} 
= 
\min_{\omega\in\Theta(\sigma,\eta)}\;\max_{i=1,\dots,|\omega|} 
H(\omega_{i-1},\omega_i) 
\;\;\; . 
\label{almin} 
\end{equation} 
We remark that the function 
$\Phi:\cc{S}\times\cc{S}\longrightarrow\bb{R}$ 
is symmetric, 
namely $\Phi(\sigma,\eta)=\Phi(\eta,\sigma)$ for any $\sigma,\eta\in\cc{S}$. 
 
We give, now, the important notion of cycle: 
we say that $\cc{A}\subset\cc{S}$ is a cycle iff for each 
$\sigma,\eta\in \cc{A}$ 
\begin{equation} 
\Phi(\sigma,\eta)< 
\min_{\zeta\in\cc{S}\setminus \cc{A}} 
\Phi(\sigma,\zeta)
\;\;\; . 
\end{equation} 
In other words starting from any configuration in the cycle $\cc{A}$, the 
energetic barrier that must be bypassed to visit any other configuration 
in $\cc{A}$ is smaller than the one  seen to exit the cycle itself. 

Given a cycle $\cc{A}\subset\cc{S}$ we 
denote by $F(\cc{A})$ the set of the minima of 
the energy in $\cc{A}$, namely 
\begin{equation} 
F(\cc{A}):=\{\sigma\in \cc{A}:\; \min_{\eta\in \cc{A}}H(\eta)=H(\sigma)\} 
\;\;\; ; 
\end{equation} 
we also write $H(F(\cc{A}))=H(\eta)$ with $\eta\in F(\cc{A})$. 
Given $\eta\in F(\cc{A})$ we define 
\begin{equation} 
\Phi(\cc{A}):=\min_{\zeta\in\cc{S}\setminus \cc{A}} \Phi(\eta,\zeta) 
\label{carat} 
\end{equation} 
(it is trivial that $\Phi(\cc{A})$ does not depend on the choice of 
$\eta\in F(\cc{A})$), 
and the set 
\begin{equation} 
U(\cc{A}):=\{\zeta\in\cc{S}\setminus\cc{A}: 
\;\exists\eta\in \cc{A} \textrm{ such that } 
H(\eta,\zeta)=\Phi(\cc{A})\} 
\;\;\; . 
\label{carat1} 
\end{equation} 

Now, for any $\eta\in\cc{S}$ let 
$\bb{P}_{\eta}$ be the probability over the process when the system is 
prepared in $\sigma_0=\eta$ and  
\begin{equation} 
\tau_{\cc{D}}:=\inf\{n\ge 0:\; \sigma_n\in \cc{D}\} 
\label{entrata} 
\end{equation} 
for any $\cc{D}\subset\cc{S}$. 
We restate, without proof, some of the results of 
\cite{[OS]} that we will use in the sequel: 
\begin{lemma} 
\label{l:enzo0} 
Given $\cc{G}\subset\cc{S}$, 
let $\sigma\in \cc{G}$ and $\sigma'\in\cc{S}\setminus \cc{G}$ 
such that: $i)$ there exist $\sigma^*\in\cc{S}\setminus \cc{G}$ and a path 
$\omega=\{\omega_0=\sigma,\dots,\omega_n=\sigma^*\}$ such that 
$\omega_i\in \cc{G}$ and 
$H(\omega_{i-1},\omega_i)<H(\omega_{n-1},\omega_n)=:\Gamma$ for 
any $i=1,\dots,n-1$; 
$ii)$ there exists a path 
$\omega'=\{\omega'_0=\sigma^*,\dots,\omega'_n=\sigma'\}$ such that 
$\omega'_i\in\cc{S}\setminus \cc{G}$ and 
$H(\omega'_{i-1},\omega'_i)<\Gamma$ for any $i=1,\dots,n$;
$iii)$ $\min_{\sigma\in\cc{G},\eta\in\cc{S}\setminus\cc{G}}H(\sigma,\eta)
        \ge\Gamma$
if and only if $\sigma=\omega_{n-1}$ and $\eta=\omega_n$.
If we define 
\begin{displaymath} 
\cc{A}:=\left\{\eta\in\cc{S}:\;\exists 
\omega=\{\omega_0=\eta,\dots,\omega_n=\sigma\}\; 
{\mathrm{such\; that}}\; \omega_1,\dots,\omega_{n-1}\in \cc{G}\; 
{\mathrm{and}}\; 
\Phi_{\omega}<\Gamma\right\} 
\end{displaymath} 
then 
$i)$ $\cc{A}\subset \cc{G}$; 
$ii)$ 
$\cc{A}$ is a cycle with $\Phi(\cc{A})=\Gamma$ and $\sigma^*\in U(\cc{A})$; 
$iii)$ 
$\Phi(\sigma,\sigma')=\Gamma$ (that is $\Gamma$ is the minmax 
between $\sigma$ and $\sigma'$). 
\end{lemma} 
 
\begin{lemma} 
\label{l:enzo} 
Given a cycle $\cc{A}\subset\cc{S}$, 
\item{$i)$} 
for all $\varepsilon>0$ and for all $\sigma\in F(\cc{A})$ 
\begin{equation} 
\bb{P}_{\sigma}\left( 
\exp\{\Phi(\cc{A})-H(F(\cc{A}))-\beta\varepsilon\}< 
\tau_{\cc{S}\setminus \cc{A}}< 
\exp\{\Phi(\cc{A})-H(F(\cc{A}))+\beta\varepsilon\}\right) 
\; 
\stackrel{\scriptstyle\beta\to\infty}{\longrightarrow}\; 
1 
\end{equation} 
\item{$ii)$} 
there exists $\delta>0$ such that for any $\sigma,\eta\in \cc{A}$ 
\begin{equation} 
\bb{P}_{\sigma}\left( 
\tau_{\eta}<\tau_{\cc{S}\setminus \cc{A}},\; 
\tau_{\eta}< 
\exp\{\Phi(\cc{A})-H(F(\cc{A}))-\beta\delta\}\right) 
\; 
\stackrel{\scriptstyle\beta\to\infty}{\longrightarrow}\; 
1 
\end{equation} 
\item{$iii)$} 
for any $\sigma\in \cc{A}$ 
\begin{equation} 
\bb{P}_{\sigma}\left( 
\sigma_{\tau_{\cc{S}\setminus \cc{A}}}\in U(\cc{A})\right) 
\; 
\stackrel{\scriptstyle\beta\to\infty}{\longrightarrow}\; 
1 
\end{equation} 
\item{$iv)$} 
for any $\sigma\in\cc{A}$, $\eta\in U(\cc{A})$, $\varepsilon>0$ and $\beta$
sufficiently large
\begin{equation}
\bb{P}_\sigma\left(\sigma_{\tau_{\cc{S}\setminus\cc{A}}}=\eta\right) 
\ge e^{-\beta\varepsilon}
\end{equation}
\end{lemma} 
\par\noindent 
We note that the lower bound on $\tau_{\cc{S}\setminus \cc{A}}$ in the 
statement $i)$ in Lemma \ref{l:enzo} 
is an easy consequence of the reversibility property: 
\begin{lemma} 
\label{l:rev} 
For any $\sigma,\eta\in\cc{S}$ such that $\Phi(\sigma,\eta)-H(\sigma)>0$ and 
for any $\delta>0$ 
\begin{equation} 
\bb{P}_{\sigma}\left(\tau_{\eta}> 
\exp\{\Phi(\sigma,\eta)-H(\sigma)-\beta\delta\}\right) 
\; 
\stackrel{\scriptstyle\beta\to\infty}{\longrightarrow}\; 
1 
\;\;\; . 
\label{rever} 
\end{equation} 
\end{lemma} 
 
Few important remarks which are very peculiar of our PCA model. 
As it has been noticed above if the system is in the state $\eta\in\cc{S}$, 
then there exists a unique configuration where it jumps with high probability. 
This configuration has been denoted by $\rmT\eta$. Thus, given 
$\eta\in\cc{S}$ we define the {\it downhill} path starting from $\eta$ 
as the unique path 
$\omega=\{\omega_0,\dots ,\omega_n\}$ such that $\omega_0=\eta$, 
$\rmT\omega_{i-1}=\omega_{i}$ for any $i=1,\dots ,n$, and 
$\omega_n$ is a trap; we also set $\widehat\eta:=\omega_n$. 
We remark that to each $\eta\in\cc{S}$ we can associate either 
a unique stable configuration or a unique stable pair.
We define the {\it basin of attraction} of a trap $\eta\in\cc{M}$ as the 
set 
\begin{equation} 
\cc{B}(\eta) := 
\{\zeta\in\cc{S}:\;\widehat\zeta=\eta\} 
\label{attr} 
\end{equation} 
and the {\it truncated} basin of attraction 
$\overline{\cc{B}}(\sigma)\subset\cc{B}(\sigma)$ 
as the set of all the configurations 
$\eta\in\cc{B}(\sigma)$ such that 
\begin{equation} 
\Phi(\eta,\sigma)<\min_{\zeta\in\cc{S}\setminus\cc{B}(\eta)} 
\Phi(\eta,\zeta) 
\;\;\; . 
\label{tron} 
\end{equation} 
It can be easily proven that $\overline{\cc{B}}(\eta)$ is a cycle. 

In the following we will often have to evaluate 
$\Upsilon(\eta):= 
\min_{\zeta\in\cc{S}\setminus\cc{B}(\eta)}\Phi(\eta,\zeta)$ for 
some trap $\eta\in\cc{M}$. A convenient way to proceed is the following: 
say that a path 
$\omega=\{\omega_0,\dots ,\omega_n\}$ is uphill iff 
the path $\omega'=\{\omega'_0,\dots,\omega'_n\}$, where 
$\omega'_i=\omega_{n-i}$ for any $i=0,\dots,n$, is downhill. 
Consider the set $\Xi(\eta)$ of paths $\omega=\{\omega_0,\dots,\omega_n\}$ 
such that 
$\omega_0=\eta$, $\{\omega_0,\dots,\omega_{n-1}\}$ 
is an uphill path in $\cc{B}(\eta)$ and 
$\omega_{n}\in\cc{S}\setminus \cc{B}(\eta)$. 
We remark that 
$\Upsilon(\eta)$, in words the barrier that must be bypassed 
to exit from the basin $\cc{B}(\eta)$, is given by
\begin{equation} 
\Upsilon(\eta)= 
\min_{\omega\in\Xi(\eta)}\Phi_{\omega} 
\;\;\; . 
\label{caut} 
\end{equation} 

\subsec{Behavior of traps}{cu:comp} 
\par\noindent 
In this subsection we 
clarify the geometrical conditions for the shrinking or the growing 
of a trap, 
that is we study the 
evolution of the system prepared in a stable configuration or in a 
stable pair. 
We let $\lcrit:=[2/h]+1$. 

We first consider the case of a single rectangular 
droplet of chessboard or pluses 
in the sea of minuses; we show that if the droplet is small enough, namely 
its shortest side is smaller than $\lcrit$, then it tends 
to shrink, otherwise it tends to grow. 
\begin{proposition} 
\label{p:contr} 
Let $\zeta\in\{\puno,\eC,\oC\}$; $\eta\in\cc{M}$ such  
that there exists a rectangle $R_{\ell,m}$, 
with $2\le\ell\le m$, such that 
$\eta_{\Lambda\setminus\core{R}_{\ell,m}}= 
 \muno_{\Lambda\setminus\core{R}_{\ell,m}}$ 
and $\eta_{\core{R}_{\ell,m}}=\zeta_{\core{R}_{\ell,m}}$. 
Thus 
\item{$i)$} 
if $\ell<\lcrit$, then $\eta$ is subcritical, that is 
$\bb{P}_{\eta}(\tau_{\muno}<\tau_{\cc{S}_{\cc{C}}\cup\cc{S}_{\puno}}) 
\; 
\stackrel{\scriptstyle\beta\to\infty}{\longrightarrow}\; 
1$, 
and for any $\varepsilon>0$ 
\begin{equation} 
\bb{P}_{\eta}\left( 
\exp\left\{2\beta h(\ell-1) -\beta\varepsilon\right\}< 
\tau_{\muno}< 
\exp\left\{2\beta h(\ell-1) +\beta\varepsilon\right\}\right) 
\; 
\stackrel{\scriptstyle\beta\to\infty}{\longrightarrow}\; 
1 
\;\;\; ; 
\end{equation} 
\item{$ii)$} 
if $\ell\ge\lcrit$, then $\eta$ is supercritical, that is 
$\bb{P}_{\eta}(\tau_{\cc{S}_{\cc{C}}\cup\cc{S}_{\puno}}<\tau_{\muno}) 
\; 
\stackrel{\scriptstyle\beta\to\infty}{\longrightarrow}\; 
1$, 
and for any $\varepsilon>0$ 
\begin{equation} 
\bb{P}_{\eta}\left( 
\exp\left\{2\beta (2-h) -\beta\varepsilon\right\}< 
\tau_{\cc{S}_{\cc{C}}\cup\cc{S}_{\puno}}< 
\exp\left\{2\beta (2-h) +\beta\varepsilon\right\}\right) 
\; 
\stackrel{\scriptstyle\beta\to\infty}{\longrightarrow}\; 
1 
\;\;\; . 
\end{equation} 
\end{proposition} 
 
Similar results can be stated in the case of 
a single droplet trap plunged inside the sea of chessboard. 
\begin{proposition} 
\label{p:conch} 
Let $C\in\cc{C}$ and 
$\eta\in\cc{M}$ a trap such that there exists a rectangle $R_{\ell,m}$, 
with $2\le\ell\le m$, and 
$\eta_{\Lambda\setminus\core{R}_{\ell,m}}=C_{\Lambda\setminus\core{R}_{\ell,m}}$ 
and $\eta_{\core{R}_{\ell,m}}=\puno_{\core{R}_{\ell,m}}$. 
Thus 
\item{$i)$} 
if $\ell<\lcrit$, then  
$\bb{P}_{\eta}(\tau_{\cc{C}}<\tau_{+\underline 1}) 
\; 
\stackrel{\scriptstyle\beta\to\infty}{\longrightarrow}\; 
1$, 
and for any $\varepsilon>0$ 
\begin{equation} 
\bb{P}_{\eta}\left( 
\exp\left\{2\beta h(\ell-1) -\beta\varepsilon\right\}< 
\tau_{\cc{C}}< 
\exp\left\{2\beta h(\ell-1) +\beta\varepsilon\right\}\right) 
\; 
\stackrel{\scriptstyle\beta\to\infty}{\longrightarrow}\; 
1 
\;\;\; ; 
\end{equation} 
\item{$ii)$} 
if $\ell\ge\lcrit$, then  
$\bb{P}_{\eta}(\tau_{+\underline 1}<\tau_{\cc{C}}) 
\; 
\stackrel{\scriptstyle\beta\to\infty}{\longrightarrow}\; 
1$, 
and for any $\varepsilon>0$ 
\begin{equation} 
\bb{P}_{\eta}\left( 
\exp\left\{2\beta (2-h) -\beta\varepsilon\right\}< 
\tau_{+\underline 1}< 
\exp\left\{2\beta (2-h) +\beta\varepsilon\right\}\right) 
\; 
\stackrel{\scriptstyle\beta\to\infty}{\longrightarrow}\; 
1 
\;\;\; . 
\end{equation} 
\end{proposition} 
 
Now, we give two heuristic arguments supporting the Propositions above. 
We consider the case $\ell=m$ even and 
$\eta_{\core{R}_{\ell,\ell}}=\oC_{\core{R}_{\ell,\ell}}$: 
by using (\ref{hl}) 
one can show that 
the energy of $\eta$, with respect to the configuration 
$\muno$, is 
\begin{equation} 
E_{\Lambda}^h(\eta)-E_{\Lambda}^h(\muno)= 
-2h\ell^2+8\ell 
\;\;\; . 
\label{enqua} 
\end{equation} 
Thus, 
$E_{\Lambda}^h(\eta)-E_{\Lambda}^h(\muno)$ 
is a parabola whose maximum is 
achieved at $\ell=2/h$ suggesting the conjecture that  the critical length 
is $2/h$. 

A dynamical argument strengthens this conjecture. Consider the most efficient 
growing mechanism: from results in 
Table \ref{tab:sin} 
this mechanism is the appearance of a single plus protuberance 
adjacent to one of the four sides of the rectangle. The probability 
associated to such an event is $\exp\{-2\beta(2-h)\}$, so that the typical time 
to see this event is $\tau_{\rm gr}\sim\exp\{2\beta(2-h)\}$. 
In Fig. \ref{f:cre} 
it is shown that once the protuberance has appeared on one the four 
sides of the rectangle, with high probability 
a new slice is filled with chessboard. 
\setlength{\unitlength}{1pt} 
\begin{figure} 
\begin{picture}(200,150)(-250,-40) 
\thinlines 
\put(-130,90){$-$}\put(-120,90){$+$} 
\put(-130,80){$+$}\put(-120,80){$-$} 
\put(-130,70){$-$}\put(-120,70){$+$} 
\put(-130,60){$+$}\put(-120,60){$-$} 
\put(-130,50){$-$}\put(-120,50){$+$} 
\put(-130,40){$+$}\put(-120,40){$-$} 
\put(-135,37){\line(1,0){25}} 
\put(-110,37){\line(0,1){63}} 
\put(-135,100){\line(1,0){25}} 
\put(-100,60){$\stackrel{\scriptstyle e^{-2\beta(2-h)}}{\longrightarrow}$} 
\put(-60,90){$+$}\put(-50,90){$-$} 
\put(-60,80){$-$}\put(-50,80){$+$}\put(-40,80){$-$} 
\put(-60,70){$+$}\put(-50,70){$-$}\put(-40,70){$+$} 
\put(-60,60){$-$}\put(-50,60){$+$}\put(-40,60){$-$} 
\put(-60,50){$+$}\put(-50,50){$-$} 
\put(-60,40){$-$}\put(-50,40){$+$} 
\put(-65,37){\line(1,0){25}} 
\put(-40,37){\line(0,1){20}} 
\put(-40,57){\line(1,0){10}} 
\put(-30,57){\line(0,1){32}} 
\put(-30,89){\line(-1,0){10}} 
\put(-40,89){\line(0,1){11}} 
\put(-65,100){\line(1,0){25}} 
\put(-25,60){$\stackrel{\scriptstyle\phantom{e^{-2\beta h}}}{\longrightarrow}$} 
\put(5,90){$-$}\put(15,90){$+$}\put(25,90){$-$} 
\put(5,80){$+$}\put(15,80){$-$}\put(25,80){$+$} 
\put(5,70){$-$}\put(15,70){$+$}\put(25,70){$-$} 
\put(5,60){$+$}\put(15,60){$-$}\put(25,60){$+$} 
\put(5,50){$-$}\put(15,50){$+$}\put(25,50){$-$} 
\put(5,40){$+$}\put(15,40){$-$} 
\put(0,37){\line(1,0){25}} 
\put(25,37){\line(0,1){10}} 
\put(25,47){\line(1,0){10}} 
\put(35,47){\line(0,1){53}} 
\put(0,100){\line(1,0){35}} 
\put(40,60){$\stackrel{\scriptstyle\phantom{e^{-2\beta h}}}{\longrightarrow}$} 
\put(70,90){$+$}\put(80,90){$-$}\put(90,90){$+$} 
\put(70,80){$-$}\put(80,80){$+$}\put(90,80){$-$} 
\put(70,70){$+$}\put(80,70){$-$}\put(90,70){$+$} 
\put(70,60){$-$}\put(80,60){$+$}\put(90,60){$-$} 
\put(70,50){$+$}\put(80,50){$-$}\put(90,50){$+$} 
\put(70,40){$-$}\put(80,40){$+$}\put(90,40){$-$} 
\put(65,37){\line(1,0){35}} 
\put(100,37){\line(0,1){63}} 
\put(65,100){\line(1,0){35}} 
\end{picture} 
\vskip -2.5 cm  
\caption[Chessboard droplet growth]{Growth of a chessboard droplet inside 
the sea of minuses: appearing of a protuberance.} 
\label{f:cre} 
\end{figure}
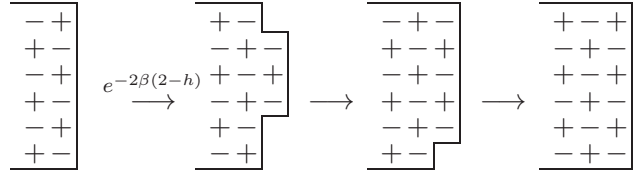 

Now we have to look for the efficient shrinking mechanism: Table 
\ref{tab:sin} suggests 
this mechanism is the ``minus corner persistence", 
that is a minus spin on one of the four corners of the rectangle is kept 
fixed during the flip--flop of the stable pair. 
By reiterating this mechanism 
$\ell-1$ times (see Fig. \ref{f:con}) 
a full slice of the droplet is erased. 
In terms of probability each 
step costs $\exp\{-2\beta h\}$, 
hence the typical 
shrinking time is $\tau_{\rm sh}\sim\exp\{2\beta h(\ell-1)\}$. By 
comparing $\tau_{\rm gr}$ and $\tau_{\rm sh}$ we find that 
the critical length should be $2/h$. 

A similar argument can be done in the case of a plus droplet: 
the growing mechanism is still the formation of a protuberance. About 
the shrinking mechanism: 
after a first step of ``corner erosion", like in the standard Glauber 
case a corner is flipped into minus, 
one minus spin appears at the corner. 
The best thing to do, as a second step, is to flip 
simultaneously both the minus spin and its adjacent plus spin. 
This event costs still $\exp\{-2\beta h\}$ and results in a shift of the 
minus ``lacuna" on the side of the rectangle. By iterating this mechanisms 
a sort of merlon is formed and a stable pair is 
reached in a typical time 
$\tau_{\rm sh}\sim\exp\{2\beta h(\ell-1)\}$. 
\setlength{\unitlength}{1pt} 
\begin{figure} 
\begin{picture}(200,150)(-190,-40) 
\thinlines 
\put(-130,90){$-$}\put(-120,90){$+$}\put(-110,90){$-$} 
\put(-130,80){$+$}\put(-120,80){$-$}\put(-110,80){$+$} 
\put(-130,70){$-$}\put(-120,70){$+$}\put(-110,70){$-$} 
\put(-130,60){$+$}\put(-120,60){$-$}\put(-110,60){$+$} 
\put(-130,50){$-$}\put(-120,50){$+$}\put(-110,50){$-$} 
\put(-130,40){$+$}\put(-120,40){$-$}\put(-110,40){$+$} 
\put(-135,37){\line(1,0){35}} 
\put(-100,37){\line(0,1){63}} 
\put(-135,100){\line(1,0){35}} 
\put(-90,60){$\stackrel{\scriptstyle e^{-2\beta h}}{\longrightarrow}$} 
\put(-60,90){$+$}\put(-50,90){$-$} 
\put(-60,80){$-$}\put(-50,80){$+$}\put(-40,80){$-$} 
\put(-60,70){$+$}\put(-50,70){$-$}\put(-40,70){$+$} 
\put(-60,60){$-$}\put(-50,60){$+$}\put(-40,60){$-$} 
\put(-60,50){$+$}\put(-50,50){$-$}\put(-40,50){$+$} 
\put(-60,40){$-$}\put(-50,40){$+$}\put(-40,40){$-$} 
\put(-65,37){\line(1,0){35}} 
\put(-30,37){\line(0,1){52}} 
\put(-30,89){\line(-1,0){10}} 
\put(-40,89){\line(0,1){11}} 
\put(-65,100){\line(1,0){25}} 
\put(-25,60){$\stackrel{\scriptstyle e^{-2\beta h}}{\longrightarrow}$} 
\put(5,90){$-$}\put(15,90){$+$} 
\put(5,80){$+$}\put(15,80){$-$} 
\put(5,70){$-$}\put(15,70){$+$}\put(25,70){$-$} 
\put(5,60){$+$}\put(15,60){$-$}\put(25,60){$+$} 
\put(5,50){$-$}\put(15,50){$+$}\put(25,50){$-$} 
\put(5,40){$+$}\put(15,40){$-$}\put(25,40){$+$} 
\put(0,37){\line(1,0){35}} 
\put(35,37){\line(0,1){42}} 
\put(35,79){\line(-1,0){10}} 
\put(25,79){\line(0,1){21}} 
\put(0,100){\line(1,0){25}} 
\put(40,60){$\stackrel{\scriptstyle e^{-2\beta h}}{\longrightarrow}$} 
\put(70,90){$+$}\put(80,90){$-$} 
\put(70,80){$-$}\put(80,80){$+$} 
\put(70,70){$+$}\put(80,70){$-$} 
\put(70,60){$-$}\put(80,60){$+$}\put(90,60){$-$} 
\put(70,50){$+$}\put(80,50){$-$}\put(90,50){$+$} 
\put(70,40){$-$}\put(80,40){$+$}\put(90,40){$-$} 
\put(65,37){\line(1,0){35}} 
\put(100,37){\line(0,1){32}} 
\put(100,69){\line(-1,0){10}} 
\put(90,69){\line(0,1){31}} 
\put(65,100){\line(1,0){25}} 
\put(105,60){$\stackrel{\scriptstyle e^{-2\beta h}}{\longrightarrow}$} 
\put(135,90){$-$}\put(145,90){$+$} 
\put(135,80){$+$}\put(145,80){$-$} 
\put(135,70){$-$}\put(145,70){$+$} 
\put(135,60){$+$}\put(145,60){$-$} 
\put(135,50){$-$}\put(145,50){$+$}\put(155,50){$-$} 
\put(135,40){$+$}\put(145,40){$-$}\put(155,40){$+$} 
\put(130,37){\line(1,0){35}} 
\put(165,37){\line(0,1){22}} 
\put(165,59){\line(-1,0){10}} 
\put(155,59){\line(0,1){41}} 
\put(130,100){\line(1,0){25}} 
\put(170,60){$\stackrel{\scriptstyle e^{-2\beta h}}{\longrightarrow}$} 
\put(200,90){$+$}\put(210,90){$-$} 
\put(200,80){$-$}\put(210,80){$+$} 
\put(200,70){$+$}\put(210,70){$-$} 
\put(200,60){$-$}\put(210,60){$+$} 
\put(200,50){$+$}\put(210,50){$-$} 
\put(200,40){$-$}\put(210,40){$+$} 
\put(195,37){\line(1,0){25}} 
\put(220,37){\line(0,1){63}} 
\put(195,100){\line(1,0){25}} 
\end{picture} 
\vskip -2.5 cm  
\caption[Chessboard droplet contraction]{Shrinking of a chessboard droplet 
inside the sea of minuses: persistence of 
a minus corner.} 
\label{f:con} 
\end{figure}
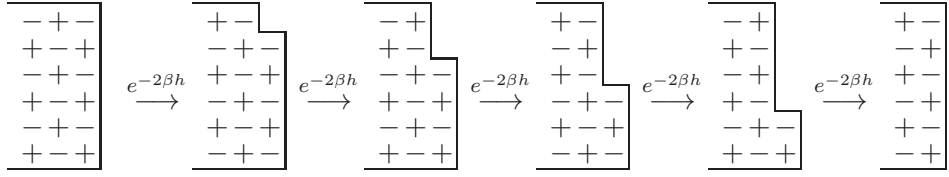 

A stronger version of the above Propositions  
can be proved; it is possible 
to describe in detail the way in which droplets shrink or grow. Indeed 
we state: 
\begin{proposition} 
\label{p:cre} 
$i)$ Let $C\in\cc{C}$, $\eta\in\cc{M}$ 
such that there exists a rectangle $R_{\ell,m}$, 
with $\ell\le m$, such that 
$\eta_{\Lambda\setminus\core{R}_{\ell,m}}= 
   \muno_{\Lambda\setminus\core{R}_{\ell,m}}$ and 
$\eta_{\core{R}_{\ell,m}}=C_{\core{R}_{\ell,m}}$. 
Let $\cc{A}'$ the set of traps $\sigma\in\cc{M}$ such that 
there exists a rectangle $R$ with side lengths $(\ell,m+1)$ or $(\ell+1,m)$ 
such that $\core{R}\supset\core{R}_{\ell,m}$, 
$\sigma_{\Lambda\setminus\core{R}}=\muno_{\Lambda\setminus\core{R}}$ and 
$\sigma_{\core{R}}=C_{\core{R}}$. 
Let $\cc{A}''$ the set of traps $\sigma\in\cc{M}$ such that 
there exists a rectangle $R$ with side lengths $(\ell,m-1)$ 
such that $\core{R}\subset\core{R}_{\ell,m}$, 
$\sigma_{\Lambda\setminus\core{R}}=\muno_{\Lambda\setminus\core{R}}$ and 
$\sigma_{\core{R}}=C_{\core{R}}$. 
If $\lcrit\le\ell$ then 
\begin{displaymath} 
\Upsilon(\eta)=H(\eta)+2\beta(2-h) 
\;\;\; and\;\;\; 
\bb{P}_{\eta}\left( 
\sigma_{\tau_{\cc{S}\setminus\cc{B}(\eta)}} 
\in\bigcup_{\sigma\in\cc{A}'}\cc{B}(\sigma)\right) 
\;\stackrel{\scriptstyle\beta\to\infty}{\longrightarrow}\; 1 
\;\;\; , 
\end{displaymath} 
that is starting from $\eta$ the system exits 
$\cc{B}(\eta)$ and enters into one of the basins $\cc{B}(\sigma)$, for 
some $\sigma\in\cc{A}'$. 
If $\ell<\lcrit$ then 
\begin{displaymath} 
\Upsilon(\eta)=H(\eta)+2\beta h(\ell-1) 
\;\;\; and\;\;\; 
\bb{P}_{\eta}\left( 
\sigma_{\tau_{\cc{S}\setminus\cc{B}(\eta)}} 
\in \cc{A}''\right) 
\;\stackrel{\scriptstyle\beta\to\infty}{\longrightarrow}\; 1 
\;\;\; , 
\end{displaymath} 
that is starting from $\eta$ the system exits 
$\cc{B}(\eta)$ and reaches directly one of the traps $\sigma$ in $\cc{A}''$. 
\par\noindent 
$ii)$ Let $C\in\cc{C}$, $\eta\in\cc{M}$ 
such that there exists a rectangle $R_{\ell,m}$, 
with $\ell\le m$, such that 
$\eta_{\Lambda\setminus\core{R}_{\ell,m}}= 
   C_{\Lambda\setminus\core{R}_{\ell,m}}$ and 
$\eta_{\core{R}_{\ell,m}}=\puno_{\core{R}_{\ell,m}}$. 
Let $\cc{A}'$ the set of traps $\sigma\in\cc{M}$ such that 
there exists a rectangle $R$ with side lengths $(\ell,m+1)$ or $(\ell+1,m)$ 
such that $\core{R}\supset\core{R}_{\ell,m}$, 
$\sigma_{\Lambda\setminus\core{R}}=C_{\Lambda\setminus\core{R}}$ and 
$\sigma_{\core{R}}=\puno_{\core{R}}$. 
Let $\cc{A}''$ the set of traps $\sigma\in\cc{M}$ such that 
there exists a rectangle $R$ with side lengths $(\ell,m-1)$ 
such that $\core{R}\subset\core{R}_{\ell,m}$, 
$\sigma_{\Lambda\setminus\core{R}}=C_{\Lambda\setminus\core{R}}$ and 
$\sigma_{\core{R}}=\puno_{\core{R}}$. 
If $\lcrit\le\ell$ then 
\begin{displaymath} 
\Upsilon(\eta)=H(\eta)+2\beta(2-h) 
\;\;\; and\;\;\; 
\bb{P}_{\eta}\left( 
\sigma_{\tau_{\cc{S}\setminus\cc{B}(\eta)}} 
\in\bigcup_{\sigma\in\cc{A}'}\cc{B}(\sigma)\right) 
\;\stackrel{\scriptstyle\beta\to\infty}{\longrightarrow}\; 1 
\;\;\; , 
\end{displaymath} 
that is starting from $\eta$ the system exits 
$\cc{B}(\eta)$ and enters into one of the basins $\cc{B}(\sigma)$, for 
some $\sigma\in\cc{A}'$. 
If $\ell<\lcrit$ then 
\begin{displaymath} 
\Upsilon(\eta)=H(\eta)+2\beta h(\ell-1) 
\;\;\; and\;\;\; 
\bb{P}_{\eta}\left( 
\sigma_{\tau_{\cc{S}\setminus\cc{B}(\eta)}} 
\in \cc{A}''\right) 
\;\stackrel{\scriptstyle\beta\to\infty}{\longrightarrow}\; 1 
\;\;\; , 
\end{displaymath} 
that is starting from $\eta$ the system exits 
$\cc{B}(\eta)$ and reaches directly one of the traps $\sigma$ in $\cc{A}''$. 
\end{proposition} 

Finally, we state under which conditions a general 
trap shrinks. A trap is made of rectangles 
of pluses or chessboard inside a minus or chessboard sea. The idea is that 
the configuration shrinks iff each single rectangular 
cluster shrinks. 
Note that the following proposition 
strictly contains Propositions \ref{p:contr} 
and \ref{p:conch}. 
\begin{proposition} 
\label{p:congl} 
$i)$ Let $C\in\cc{C}$ and $\eta\in\cc{M}_C$. There exist 
$k\ge 1$ pairwise non--interacting rectangles 
$R_{\ell_1,m_1},\dots,R_{\ell_k,m_k}$ such that $2\le\ell_i\le m_i\le L-2$ for 
any $i=1,\dots,k$, $\eta_{\Re}=\puno_{\Re}$ and 
$\eta_{\Lambda\setminus\Re}=C_{\Lambda\setminus\Re}$ where 
$\Re:=\bigcup_{i=1}^k\core{R}_{\ell_i,m_i}$. 
Thus, 
if $\ell_i<\lcrit$ for any $i=1,\dots,k$, then  
$\bb{P}_{\eta}(\tau_{\cc{C}}<\tau_{+\underline 1}) 
\; 
\stackrel{\scriptstyle\beta\to\infty}{\longrightarrow}\; 
1$, 
and for any $\varepsilon>0$ 
\begin{equation} 
\bb{P}_{\eta}\left( 
\exp\left\{2\beta h(\ell-1) -\beta\varepsilon\right\}< 
\tau_{\cc{C}}< 
\exp\left\{2\beta h(\ell-1) +\beta\varepsilon\right\}\right) 
\; 
\stackrel{\scriptstyle\beta\to\infty}{\longrightarrow}\; 
1 
\;\;\; , 
\end{equation} 
where $\ell:=\max\{\ell_1,\dots,\ell_k\}$. 
If there exists $j\in\{1,\dots,k\}$ such that 
$\ell_j\ge\lcrit$, then  
$\bb{P}_{\eta}(\tau_{+\underline 1}<\tau_{\cc{C}}) 
\; 
\stackrel{\scriptstyle\beta\to\infty}{\longrightarrow}\; 
1$, 
and for any $\varepsilon>0$ 
\begin{equation} 
\bb{P}_{\eta}\left( 
\exp\left\{2\beta (2-h) -\beta\varepsilon\right\}< 
\tau_{+\underline 1}< 
\exp\left\{2\beta (2-h) +\beta\varepsilon\right\}\right) 
\; 
\stackrel{\scriptstyle\beta\to\infty}{\longrightarrow}\; 
1 
\end{equation} 
(note that in the case $k=1$ we recover Proposition \ref{p:conch}). 
$ii)$ 
We consider, now, a situation where in a sea of minuses there are 
rectangles of chessboard and, possibly, rectangles of pluses 
inside the sea of minuses or inside the chessboard droplets. More 
precisely,
let $\eta\in\cc{M}$. We suppose that there exist $k\ge 1$ 
rectangles 
$R_{\ell_1,m_1},\dots,R_{\ell_k,m_k}$, with $2\le\ell_i\le m_i\le L-2$ for 
any $i=1,\dots,k$, and there exists an integer $s\in\{1,\dots,k\}$ such 
that the conditions of point $(iii)$ in Proposition \ref{p:res} are 
satisfied (note that in the case $s=0$ the trap is a stable 
configuration; in the case $s=0$ and $k=1$ Proposition \ref{p:contr} is 
recovered). 
Thus, 
if $\ell_i<\lcrit$ for any $i=1,\dots,k$, then  
$\bb{P}_{\eta}(\tau_{\muno}<\tau_{\cc{S}_{\cc{C}}\cup\cc{S}_{\puno}}) 
\; 
\stackrel{\scriptstyle\beta\to\infty}{\longrightarrow}\; 
1$, 
and for any $\varepsilon>0$ 
\begin{equation} 
\bb{P}_{\eta}\left( 
\exp\left\{2\beta h(\ell-1) -\beta\varepsilon\right\}< 
\tau_{\muno}< 
\exp\left\{2\beta h(\ell-1) +\beta\varepsilon\right\}\right) 
\; 
\stackrel{\scriptstyle\beta\to\infty}{\longrightarrow}\; 
1 
\;\;\; , 
\end{equation} 
where $\ell:=\max\{\ell_1,\dots,\ell_k\}$. 
If there exists $j\in\{1,\dots,k\}$ such that 
$\ell_j\ge\lcrit$, then $\eta$ is supercritical, that is 
$\bb{P}_{\eta}(\tau_{\cc{S}\setminus\cc{S}_{\muno}}<\tau_{\muno}) 
\; 
\stackrel{\scriptstyle\beta\to\infty}{\longrightarrow}\; 
1$, 
and for any $\varepsilon>0$ 
\begin{equation} 
\bb{P}_{\eta}\left( 
\exp\left\{2\beta (2-h) -\beta\varepsilon\right\}< 
\tau_{\cc{S}\setminus\cc{S}_{\muno}}< 
\exp\left\{2\beta (2-h) +\beta\varepsilon\right\}\right) 
\; 
\stackrel{\scriptstyle\beta\to\infty}{\longrightarrow}\; 
1 
\;\;\; . 
\end{equation} 
\end{proposition} 
 
\subsec{Exit from the metastable phase}{su:uscita} 
\par\noindent 
We can now give the theorem describing the exit from the 
metastable state. 
Suppose that the system is prepared in the metastable state, 
$\sigma_0=-{\underline 1}$, in the following theorem we state that 
the first exit time $\tau_{+{\underline 1}}$ is exponentially large in 
$\beta$ and we find its order of magnitude. 
Moreover, we state that before reaching $\puno$ the system 
visits $\cc{C}=\{\oC,\eC\}$ and that the typical time to jump from 
$\muno$ to the chessboards is the same as the time needed to jump 
from the chessboards to $\puno$. 
More precisely, 
let us denote by $\cc{Q}_{\muno}\subset\cc{S}$ the set of configurations 
$\eta\in\cc{S}$ such that there exists a rectangle 
$R_{\lcrit,\lcrit}$ such that 
$\eta_{\Lambda\setminus\core{R}_{\lcrit,\lcrit}}= 
\muno_{\Lambda\setminus\core{R}_{\lcrit,\lcrit}}$ and 
$\eta_{\core{R}_{\lcrit,\lcrit}}=C_{\core{R}_{\lcrit,\lcrit}}$ 
with $C\in\cc{C}$. 
Let $\eta\in\cc{Q}_{\muno}$, we call ``protocritical height" the energy 
\begin{equation} 
\Gamma:= 
E_{\Lambda}^h(\eta)- 
E_{\Lambda}^h(\muno)+2h(\lcrit-1)= 
-2h{\lcrit}^2 +2\lcrit(4+h)-2h 
\;\;\; , 
\label{prot-h} 
\end{equation} 
where the second equality follows from (\ref{enqua}). 
In some sense $\beta\Gamma$ is the communication height between the largest 
subcritical droplet and the smallest supercritical droplet; more precisely: 
let $\cc{O}_{\muno}\subset\cc{S}$ 
the set of configurations $\eta\in\cc{S}_{\muno}$ such that 
there exists a rectangle 
$R_{\lcrit-1,\lcrit}$ such that 
$\eta_{\Lambda\setminus\core{R}_{\lcrit-1,\lcrit}}= 
\muno_{\Lambda\setminus\core{R}_{\lcrit-1,\lcrit}}$ and 
$\eta_{\core{R}_{\lcrit-1,\lcrit}}=C_{\core{R}_{\lcrit-1,\lcrit}}$ 
with $C\in\cc{C}$. 

Let $\eta\in \cc{O}_{\muno}$: we call {\it protocritical} 
droplet corresponding to $\eta$ one of the configurations  
obtained by flipping in $\eta$ 
a minus spin external to $R_{\lcrit-1,\lcrit}$ and adjacent to one of the 
plus spins of the internal chessboard and all the spins 
associated to sites inside $R_{\lcrit-1,\lcrit}$. 
We let 
$\pi_{\muno}(\eta)$ the set of protocritical droplets corresponding 
to $\eta$ and 
$\cc{P}_{\muno}:=\cup_{\eta\in\cc{O}_{\muno}}\pi_{\muno}(\eta)$, 
the collection of the protocritical droplets. 
It is easy to check that 
$H(\eta,\zeta)=\Gamma$ for any $\eta\in\cc{O}_{\muno}$
and $\zeta\in\pi_{\muno}(\eta)$. 
\begin{theorem} 
\label{t:uscita} 
With the notation introduced above: 
$i)$ the system visits $\cc{P}_{\muno}$ before visiting $\cc{C}$, 
namely 
\begin{displaymath} 
\bb{P}_{\muno}(\tau_{\cc{P}_{\muno}}<\tau_{\cc{C}}) 
\; 
\stackrel{\scriptstyle\beta\to\infty}{\longrightarrow}\; 
1\;\;\; ; 
\end{displaymath} 
\par\noindent 
$ii)$ the system visits $\cc{C}$ before visiting $\puno$, namely 
\begin{displaymath} 
\bb{P}_{\muno}(\tau_{\cc{C}}<\tau_{\puno}) 
\; 
\stackrel{\scriptstyle\beta\to\infty}{\longrightarrow}\; 
1\;\;\; ; 
\end{displaymath} 
\par\noindent 
$iii)$ for any $\varepsilon>0$ 
\begin{displaymath} 
\bb{P}_{\muno}( 
e^{\beta\Gamma-\beta\varepsilon} 
<\tau_{\cc{C}}< 
e^{\beta\Gamma+\beta\varepsilon} 
) 
\; 
\stackrel{\scriptstyle\beta\to\infty}{\longrightarrow}\; 
1\;\;\; ; 
\end{displaymath} 
\par\noindent 
$iv)$ for any $\varepsilon>0$ 
\begin{displaymath} 
\bb{P}_{\muno}( 
e^{\beta\Gamma-\beta\varepsilon} 
<\tau_{\puno}< 
e^{\beta\Gamma+\beta\varepsilon} 
) 
\; 
\stackrel{\scriptstyle\beta\to\infty}{\longrightarrow}\; 
1\;\;\; . 
\end{displaymath} 
\end{theorem} 
The proof of Theorem \ref{t:uscita} will be the argument of 
Section \ref{s:dim}.

In the above theorem we have stated that during the exit from the 
metastable $\muno$ state, the system visits the competing metastable state 
$\cc{C}$ and, finally, reaches the stable state $\puno$. 
Now, we want to give a more precise description of the path followed 
by the system during its exit from $\muno$; first of all 
we define a suitable tube $\cc{T}_{\muno}$.
Let $\cc{O}_{\muno}^{(0)}:=\cc{O}_{\muno}$  
the set of $\lcrit\times(\lcrit-1)$ 
chessboard droplets in the sea of minuses. 
For each $k=0,1,\dots,2\lcrit-6$ we define recursively the sets 
$\cc{O}_{\muno}^{(k)}(\eta^{(k-1)})$, where 
$\eta^{(k-1)}\in \cc{O}_{\muno}^{(k-1)}(\eta^{(k-2)})$:  
let $\eta^{(k-1)}\in \cc{O}_{\muno}^{(k-1)}(\eta^{(k-2)})$ a  
configuration such that there exists a rectangle $R_{\ell,m}$, with  
$2\le\ell\le m$, such that  
$\eta^{(k-1)}_{\core{R}_{\ell,m}}=C_{\core{R}_{\ell,m}}$, with $C\in\cc{C}$, 
and  
$\eta^{(k-1)}_{\Lambda\setminus\core{R}_{\ell,m}}= 
 \muno_{\Lambda\setminus\core{R}_{\ell,m}}$.  
Then we define  
$\cc{O}_{\muno}^{(k)}(\eta^{(k-1)})$ 
as the collection of configurations $\zeta$ such that 
there exists a rectangle $R_{\ell,m-1}$ such that 
$\core{R}_{\ell,m-1}\subset\core{R}_{\ell,m}$, 
$\zeta_{\Lambda\setminus\core{R}_{\ell,m-1}}= 
 \muno_{\Lambda\setminus\core{R}_{\ell,m-1}}$ and 
$\zeta_{\core{R}_{\ell,m-1}}=C_{\core{R}_{\ell,m-1}}$, with $C\in\cc{C}$. 
We remark that 
$\cc{O}_{\muno}^{(2\lcrit-7)}(\eta^{(2\lcrit-8)})$ 
is a set of $2\times 2$ chessboard droplets and 
$\cc{O}_{\muno}^{(2\lcrit-6)}(\eta^{(2\lcrit-7)})$ is 
made of two configurations with a plus spin in the sea of minuses. 
 
We note that Proposition \ref{p:cre} implies that 
the process enters, with high probability in the limit $\beta\to\infty$,  
into the set $\cc{O}_{\muno}^{(i+1)}(\eta^{(i)})$,  
when it exits from the basin of 
attraction $\cc{B}(\eta_i)$ with 
$\eta_i\in \cc{O}_{\muno}^{(i)}(\eta^{(i-1)})$. 

Now, given the $2\lcrit-5$ configurations 
$\eta_i\in \cc{O}_{\muno}^{(i)}(\eta^{(i-1)})$ for any   
$i=0,1,\dots,2\lcrit-6$ and the $2\lcrit-6$ integer numbers 
$t_1<\cdots<t_{2\lcrit-6}$ we set $t_0=0$ and 
$t_{2\lcrit-5}=t_{2\lcrit-6}+1$, and we say that a path 
$\omega=\{\omega_{t_0},\dots,\omega_{t_{2\lcrit-5}}\}$ belongs 
to the set 
$\cc{T}\left(\eta^{(0)},\dots,\eta^{(2\lcrit-6)};  
  t_1,\dots,t_{2\lcrit-6}\right)$ 
iff $\omega_{t_i}=\eta^{(i)}$ and 
$\omega_{t_i},\dots,\omega_{t_{i+1}-1}\in\overline{\cc{B}}(\eta^{(i)})$  
for any $i=0,\dots,2\lcrit-6$ and $\omega_{t_{2\lcrit-5}}=\muno$. 
Note that a path in 
$\cc{T}\left(\eta^{(0)},\dots,\eta^{(2\lcrit-6)};  
  t_1,\dots,t_{2\lcrit-6}\right)$ 
is one of the ``standard" shrinking paths  
that the system follows when a chessboard 
droplet $\eta^{(0)}\in \cc{O}_{\muno}^{(0)}$ shrinks.  
More precisely, given $\eta^{(0)}\in\cc{O}_{\muno}^{(0)}$,  
we define the tube  
\begin{equation} 
\label{tubcon} 
\cc{T}_{\eta^{(0)}}:= 
\bigcup_{i=1}^{2\lcrit-6} 
\bigcup_{\eta^{(i)}\in\cc{O}^{(i)}(\eta^{(i-1)})} 
\bigcup_{t_1<\cdots<t_{2\lcrit-6}} 
\cc{T}\left(\eta^{(0)},\dots,\eta^{(2\lcrit-6)};  
  t_1,\dots,t_{2\lcrit-6}\right) 
\;\;\; .  
\end{equation} 
In other words $\cc{T}_{\eta^{(0)}}$ is defined as the set of paths 
$\omega=\in\Theta(\eta^{(0)},\muno)$ such that 
there exist 
$2\lcrit-6$ configurations 
$\eta^{(i)}\in \cc{O}_{\muno}^{(i)}(\eta^{(i-1)})$ with 
$i=1,\dots,2\lcrit-6$ and $2\lcrit-6$ integer numbers 
$t_1<\cdots<t_{2\lcrit-6}$ such that 
$\omega\in 
\cc{T}\left(\eta^{(0)},\dots,\eta^{(2\lcrit-6)};  
  t_1,\dots,t_{2\lcrit-6}\right)$. 
We state, now, the following Lemma: 
\begin{lemma} 
\label{l:dis} 
Let $\eta_0\in \cc{O}_{\muno}$, we have 
\begin{displaymath} 
\bb{P}_{\eta^{(0)}}\left( 
\textrm{the trajectory } 
\left\{\sigma_0,\dots,\sigma_{\tau_{\muno}}\right\}\in\cc{T}_{\eta^{(0)}} 
\right) 
\;\stackrel{\scriptstyle\beta\to\infty}{\longrightarrow}\; 1 
\;\;\; . 
\end{displaymath} 
\end{lemma} 
 
\addcontentsline{toc}{subsection}{{\it Proof of Lemma  
\ref{l:dis}}} 
\medskip\noindent{\it Proof of Lemma \ref{l:dis}.}\/ 
The Lemma easily follows by applying recursively the  
Proposition \ref{p:cre} and the Markov property. 
\qed\smallskip 
 
Finally we define the exit tube $\cc{T}_{\muno}$: a path 
$\omega=\{\omega_0,\dots,\omega_n\}$ is an element of $\cc{T}_{\muno}$ 
iff there exist 
$2\lcrit-5$ configurations 
$\eta^{(i)}\in \cc{O}_{\muno}^{(i)}(\eta^{(i-1)})$ with 
$i=0,1,\dots,2\lcrit-6$, the integer numbers 
$t_1<\cdots<t_{2\lcrit-6}=n-1$ and 
a path 
$\omega'=\{\omega'_0,\dots,\omega'_n\}\in 
\cc{T}\left(\eta^{(0)},\dots,\eta^{(2\lcrit-6)};  
   t_0,\dots,t_{2\lcrit-6}\right)$ 
such that 
$\omega_i=\omega'_{n-1}$ for any $i=0,\dots,n$. In other words 
$\cc{T}_{\muno}$ is the set of paths obtained by time reversing one of the 
standard shrinking paths associated to the droplets in $\cc{O}_{\muno}$. 

\begin{theorem} 
\label{t:tubo} 
Let $\sigma_t$ be the process started at $\muno$, let 
$\bar{\tau}_{\muno}:=\max\{t<\tau_{\cc{S}\setminus\cc{A}_{\muno}}: 
\; \sigma_t=\muno\}$, then 
\begin{displaymath} 
\bb{P}_{\muno}\left( 
\sigma_{\tau_{\cc{S}\setminus\cc{A}_{\muno}}}\in\cc{P}_{\muno},\; 
{\mathrm{the\; trajectory}}\left\{ 
\sigma_{\bar{\tau}_{\muno}}, 
\sigma_{\bar{\tau}_{\muno}+1}, 
\dots, 
\sigma_{\tau_{\cc{S}\setminus\cc{A}_{\muno}}-1} 
\right\} 
\in\cc{T}_{\muno} 
\right) 
\;\stackrel{\scriptstyle\beta\to\infty}{\longrightarrow}\; 1 
\;\;\; . 
\end{displaymath} 
\end{theorem} 

\addcontentsline{toc}{subsection}{{\it Proof of Theorem 
\ref{t:tubo}}} 
\smallskip\noindent{\it Proof of Theorem \ref{t:tubo}.}\/ 
The theorem is a straightforward consequence of the time--reversing argument 
(see \cite{[OS],[S]}) and Lemma \ref{l:dis}. 
\qed\smallskip 
 
\sezione{The minmax between the metastable and the stable state}{s:dim} 
\par\noindent 
The proof of the theorems describing the exit of the system from the 
metastable state is based on the general lemmata given in the 
Subsection \ref{su:strum}. The highly not trivial model dependent part 
consists in finding the minmax between the metastable and the stable phases. 
It is clear that new ideas must be used to answer this question in the 
case of a parallel dynamics with respect to the Glauber case. 
Indeed, the fact that the system can jump from any configuration to any 
other, highly complicates the structure of the possible trajectories 
in the configuration space. 
 
First of all we define a sort of generalized basin of attraction of $\muno$: 
let $\cc{G}_{\muno}\subset\cc{S}_{\muno}$ 
the set 
\begin{equation} 
\cc{G}_{\muno}:=\left\{\sigma\in\cc{S}_{\muno}:\;\widehat\sigma=\muno 
\;{\mathrm{or}}\;\widehat\sigma\;{\mathrm{subcritical}}\right\} 
\;\;\; , 
\label{defes} 
\end{equation} 
where $\widehat\sigma$ subcritical means that $\widehat\sigma$ is 
a trap such that 
there exist $k\ge 1$ rectangles on the dual lattice 
$R_{\ell_1,m_1},\dots,R_{\ell_k,m_k}$, with $2\le\ell_i\le m_i\le L-2$ for 
any $i=1,\dots,k$, there exists $s\in\{0,\dots,k\}$ such 
that the conditions of point $iii)$ in Proposition \ref{p:res} are 
satisfied (note that in the case $s=0$ the trap is a stable 
configuration) and $\ell_i<\lcrit$ for any $i=1,\dots,k$. 
To fix the ideas: if $\widehat\sigma$ consisted of a single chessboard 
rectangle in the sea of minuses, then its shortest side length $\ell$ 
should be smaller than $\lcrit$. 
The set $\cc{G}_{\muno}$ is a sort of ``generalized" basin 
of attraction of the state $\muno$, in the sense that for any 
$\sigma\in\ \cc{G}_{\muno}$ the process started at $\sigma$ would visit 
$\muno$ before exiting $\cc{G}_{\muno}$ with high probability in the zero 
temperature limit, namely 
\begin{equation} 
\bb{P}_{\sigma}\left(\tau_{\muno}<\tau_{\cc{S}\setminus \cc{G}_{\muno}}\right) 
\;\stackrel{\scriptstyle\beta\to\infty}{\longrightarrow}\; 1 
\;\;\; . 
\label{pres} 
\end{equation} 
In the following Lemma we state the main properties of the basin 
$\cc{G}_{\muno}$: 
$\Gamma$ is 
the minimal energy barrier that must be bypassed to exit $\cc{G}_{\muno}$; 
a minimal exit path from $\cc{G}_{\muno}$ reaches 
$\cc{S}\setminus\cc{G}_{\muno}$ in a protocritical droplet 
$\eta\in\cc{P}_{\muno}$. 
\begin{lemma} 
\label{l:caG1} 
Let $\eta\in\cc{P}_{\muno}$, 
$i)$ there exists a path 
$\omega=\{\omega_0=\muno,\dots,\omega_n=\eta\}$ such that 
$\omega_i\in\cc{G}_{\muno}$ and 
$H(\omega_{i-1},\omega_i)<H(\omega_{n-1},\omega_n)=H(\muno)+\beta\Gamma$ 
for 
any $i=1,\dots,n-1$; 
$ii)$ there exists a path 
$\omega'=\{\omega'_0=\eta,\dots,\omega'_n\in\cc{C}\}$ such that 
$\omega'_i\in\cc{S}\setminus \cc{G}_{\muno}$ and 
$H(\omega_{i-1},\omega_i)<H(\muno)+\beta\Gamma$ for 
any $i=1,\dots,n$. 
$iii)$ $\Phi(\cc{G}_{\muno})=H(\muno)+\beta\Gamma$;
$iv)$ for all $\sigma\in\cc{G}_{\muno}$ and 
$\eta\in\cc{S}\setminus\cc{G}_{\muno}$, 
$H(\sigma,\eta)=\beta\Gamma+H(\muno)$ if and only if 
$\sigma\in\cc{O}_{\muno}$ and $\eta\in\pi_{\muno}(\sigma)$.  
\end{lemma} 
\par\noindent 
We postpone the proof of the above lemma to the end of this section. 
Let us define the set 
\begin{displaymath} 
\begin{array}{rl} 
\cc{A}_{\muno}:=\left\{\eta\in\cc{S}:\right.&\exists 
\omega=\{\omega_0=\eta,\dots,\omega_n=\muno\} 
\;{\mathrm{such\; that}}\; \\ 
&\omega_1,\dots,\omega_{n-1}\in \cc{G}_{\muno}\; 
{\mathrm{and}}\; 
\Phi_{\omega}<H(\muno)+\left.\beta\Gamma\right\} 
\end{array} 
\end{displaymath} 
From Lemma \ref{l:enzo0} and Lemma \ref{l:caG1} we have that 
$\cc{A}_{\muno}$ is a cycle, 
$\cc{A}_{\muno}\subset\cc{G}_{\muno}$,
$\Phi(\cc{A}_{\muno})=H(\muno)+\beta\Gamma$, 
$U(\cc{A}_{\muno})\supset\cc{P}_{\muno}$ 
and $\Phi(\muno,C)=\beta\Gamma+H(\muno)$,  
with $C\in\cc{C}$ (that is $\beta\Gamma+H(\muno)$ is the minmax 
between $\muno$ and $C$). 

\addcontentsline{toc}{subsection}{{\it Proof of Theorem 
\ref{t:uscita}}} 
\medskip\noindent{\it Proof of Theorem \ref{t:uscita}.}\/ 
Let $\sigma_t$ be the process started at $\muno$. We first 
try to describe the exit from the basin $\cc{G}_{\muno}$
by applying Lemma \ref{l:enzo} and using recurrence in $\cc{G}_{\muno}$.
We firstly remark that from the reversibility Lemma we have
that for each $\varepsilon>0$
\begin{equation} 
\bb{P}_{\muno}\left( 
\tau_{\cc{S}\setminus \cc{G}_{\muno}}>
e^{\beta\Gamma-\beta\varepsilon}
\right) 
\;\stackrel{\scriptstyle\beta\to\infty}{\longrightarrow}\; 1 
\label{ques} 
\end{equation} 
Now, from item $iv)$ in Lemma~\ref{l:enzo}, from the definition 
of $\cc{G}_{\muno}$ and from Proposition~\ref{p:cre} we have that for each 
$\sigma\in\cc{G}_{\muno}$ and each $\varepsilon,\delta>0$
\begin{equation}
\label{tot}
\bb{P}_\sigma\left(\exists t<e^{\beta\Gamma+\beta\delta},\,
                   \sigma_t\in\cc{S}\setminus\cc{G}_{\muno}\right) 
\ge e^{-\beta\varepsilon}
\end{equation}
For any $\varepsilon>0$, 
we set $T(\varepsilon):=\exp\{\beta\Gamma+\beta\varepsilon\}$,
$N(\varepsilon)=[\exp\{\beta\varepsilon/2\}]-1\in\bb{N}$, and consider the 
intervals
$I_k(\varepsilon):=T(\varepsilon)\exp\{-\beta\varepsilon/2\}[k,k+1)$ for any 
$k=0,\dots,N(\varepsilon)$. Then
\begin{equation}
\label{tat}
\begin{array}{rcl}
\bb{P}_{\muno}\left( 
 \tau_{\cc{S}\setminus \cc{G}_{\muno}}<e^{\beta\Gamma+\beta\varepsilon}\right) 
& = &
1-\bb{P}_{\muno}\left( 
 \tau_{\cc{S}\setminus \cc{G}_{\muno}}>e^{\beta\Gamma+\beta\varepsilon}\right) 
\\
&=&
{\displaystyle
1-\prod_{k=0}^{N(\varepsilon)}
\bb{P}_{\muno}\left( 
\sigma_t\in\cc{G}_{\muno}\,\textrm{for all}\, t\in I_k(\varepsilon)\right)}
\\
&=&
{\displaystyle
1-\sup_{\eta\in\cc{G}_{\muno}}\left[
        1-\bb{P}_{\eta}\left(\exists t<e^{\beta\Gamma+\beta\varepsilon/2},\,
          \sigma_t\in\cc{S}\setminus\cc{G}_{\muno}\right)
   \right]^{N(\varepsilon)}}\\
&\ge&
{\displaystyle
1-\left[
       1-e^{-\beta\varepsilon/2}\right]^{N(\varepsilon)}}
\end{array}
\end{equation}
where we have used (\ref{tot}).
{}From (\ref{tat}) we get the upper bound on the exit time 
\begin{equation}
\label{tit}
\bb{P}_{\muno}\left( 
 \tau_{\cc{S}\setminus \cc{G}_{\muno}}<e^{\beta\Gamma+\beta\varepsilon}\right) 
\;\stackrel{\scriptstyle\beta\to\infty}{\longrightarrow}\; 1 
\end{equation}
for any $\varepsilon>0$.
Now, by using the reversibility Lemma we have  
that 
\begin{equation} 
\bb{P}_{\muno}\left( 
\sigma_{\tau_{\cc{S}\setminus \cc{G}_{\muno}}}\in\cc{P}_{\muno}\right) 
\;\stackrel{\scriptstyle\beta\to\infty}{\longrightarrow}\; 1 
\;\;\; 
\label{doves} 
\end{equation} 
and recalling (\ref{ques}) and (\ref{tit}) we ca state that
the system prepared in $\muno$ exits $\cc{G}_{\muno}$ 
through $\cc{P}_{\muno}$ in a typical time $\exp\{\beta\Gamma\}$. 
Moreover,  
$\cc{C}\subset\cc{S}\setminus\cc{G}_{\muno}$ (the chessboards do not belong 
to $\cc{G}_{\muno}$) and equation (\ref{doves}) imply that 
$\cc{P}_{\muno}$ is visited before $\cc{C}$, more precisely 
\begin{displaymath} 
\bb{P}_{\muno}\left( 
\tau_{\cc{P}_{\muno}}< 
\tau_{\cc{C}}\right) 
\;\stackrel{\scriptstyle\beta\to\infty}{\longrightarrow}\; 1 
\;\;\; , 
\end{displaymath} 
completing the proof of the statement $i)$ in Theorem \ref{t:uscita}. 

Now, we use the Markov property to restart the system in some  
configuration of $\cc{P}_{\muno}$. 
Point $i)$ in Proposition \ref{p:cre} directly implies point  
$ii)$ in Theorem \ref{t:uscita}. Moreover, point $iii)$ in Theorem  
\ref{t:uscita} is easily proven by remarking that 
$\beta\Gamma>2\beta(2-h)$ and by using Proposition \ref{p:cre}. 

Up to now we have described the jump from $\muno$ to the chessboards. Now 
we use the Markov property to restart the system in 
$C\in\cc{C}$ and 
we prove point $iv)$ in Theorem \ref{t:uscita} by following the 
same scheme used above. We just sketch the proof: let 
$\cc{G}_{\cc{C}}\subset\cc{S}_{\cc{C}}$ be the set 
\begin{equation} 
\cc{G}_{\cc{C}}:=\left\{\sigma\in\cc{S}_{\cc{C}}:\;\widehat\sigma\in\cc{C} 
\;{\mathrm{or}}\;\widehat\sigma\;{\mathrm{subcritical}}\right\} 
\;\;\; , 
\label{defesch} 
\end{equation} 
where $\widehat\sigma$ subcritical means that $\widehat\sigma$ is 
a trap such that 
$k\ge 1$ pairwise non--interacting rectangles 
$R_{\ell_1,m_1},\dots,R_{\ell_k,m_k}$ such that 
$2\le\ell_i\le m_i\le L-2$ and $\ell_i<\lcrit$ for 
any $i=1,\dots,k$, 
$\eta_{\Re}=\puno_{\Re}$ and 
$\eta_{\Lambda\setminus\Re}=C_{\Lambda\setminus\Re}$ where 
$\Re:=\bigcup_{i=1}^k\core{R}_{\ell_i,m_i}$ and $C\in\cc{C}$. 
To fix the ideas: if $\widehat\sigma$ consisted of a single plus 
rectangle in the sea of chessboard, then its shortest side length $\ell$ 
should be smaller than $\lcrit$. 
Then we state the analogous of Lemma \ref{l:caG1}: 
let us denote by $\cc{Q}_{\cc{C}}$ the set of configurations 
$\eta\in\cc{S}$ such that there exists a rectangle 
$R_{\lcrit,\lcrit}$ such that 
$\eta_{\Lambda\setminus\core{R}_{\lcrit,\lcrit}}= 
C_{\Lambda\setminus\core{R}_{\lcrit,\lcrit}}$ and 
$\eta_{\core{R}_{\lcrit,\lcrit}}=\puno_{\core{R}_{\lcrit,\lcrit}}$ 
with $C\in\cc{C}$. 
We denote by $\cc{P}_{\cc{C}}$ the set of 
configurations $\eta\in\cc{S}$  
such that there exist a rectangle $R_{\lcrit-1,\lcrit}$, a  
site $x\in\Lambda\setminus\core{R}_{\lcrit-1,\lcrit}$, adjacent to one 
of the two sides of $\core{R}_{\lcrit-1,\lcrit}$ of length $\lcrit$, and  
$C\in\cc{C}$ such that 
$C(x)=-1$, 
$\eta_{\Lambda\setminus(\core{R}_{\lcrit,\lcrit}\cup\{x\})}= 
  C_{\Lambda\setminus(\core{R}_{\lcrit,\lcrit}\cup\{x\})}$, 
$\eta_{\core{R}_{\lcrit,\lcrit}}=\puno_{\core{R}_{\lcrit,\lcrit}}$ 
and $\eta(x)=+1$. 
\begin{lemma} 
\label{l:caGch} 
Let $C\in\cc{C}$, $\eta\in\cc{P}_{\cc{C}}$, 
$i)$ there exists a path 
$\omega=\{\omega_0=C,\omega_1,\dots,\omega_n=\eta\}$ such that 
$\omega_i\in \cc{G}_{\cc{C}}$ and 
$H(\omega_{i-1},\omega_i)<H(\omega_{n-1},\omega_n)=H(C)+\beta\Gamma$ for 
any $i=1,\dots,n-1$; 
$ii)$ there exists a path 
$\omega'=\{\omega'_0=\eta,\dots,\omega'_n=\puno\}$ such that 
$\omega'_i\in\cc{S}\setminus \cc{G}_{\cc{C}}$ and 
$H(\omega_{i-1},\omega_i)<H(C)+\beta\Gamma$ for 
any $i=1,\dots,n$. 
$iii)$ $\Phi(\cc{G}_{\cc{C}})=H(C)+\beta\Gamma$.
\end{lemma} 
\par\noindent 
As before the statement $iv)$ in Theorem \ref{t:uscita} follows from 
the Lemmata \ref{l:caGch}, \ref{l:enzo0}, \ref{l:enzo} and   
point $ii)$ in Proposition \ref{p:cre}. 
\qed\smallskip 
 
\addcontentsline{toc}{subsection}{{\it Proof of Lemma 
\ref{l:caG1}}} 
\medskip\noindent{\it Proof of Lemma \ref{l:caG1}.}\/ 
We start by proving point $i)$: 
let us consider a protocritical droplet $\eta\in\cc{P}_{\muno}$ and the 
$\lcrit\times(\lcrit-1)$ chessboard droplet $\eta^{(0)}\in\cc{O}_{\muno}$  
such that $\pi_{\muno}(\eta^{(0)})=\eta$. First of all 
we note that: $H(\eta,\eta^{(0)})=H(\muno)+\beta\Gamma$. 

Now, recall $\cc{O}_{\muno}^{(0)}=\cc{O}_{\muno}$ and 
consider a sequence of configurations  
$\eta^{(1)}\in \cc{O}_{\muno}^{(1)}(\eta^{(0)})$,$\dots$, 
$\eta_{2\lcrit-6}\in\cc{O}_{\muno}^{(2\lcrit-6)} 
     (\eta^{(2\lcrit-7)})$. From Proposition 
\ref{p:cre} we have that for any $i=1,\dots,2\lcrit-7$ the barrier 
to exit the related basin of attraction is 
\begin{displaymath} 
\begin{array}{rl} 
\Upsilon(\eta^{(i)})\le 
& 
H(\eta^{(i)})+2\beta h(\lcrit-2)<H(\eta^{(0)})+2\beta h(\lcrit-2)\\ 
&\\ 
<& 
H(\eta^{(0)})+2\beta(2-h)=H(\muno)+\beta\Gamma\\ 
\end{array} 
\;\;\; . 
\end{displaymath} 
Note that for $i=0,1$ the first inequality is, indeed, an equality.

The above inequalities allow to construct a path 
$\{\omega_0=\eta^{(0)},\dots,\omega_n=\muno\}\in\cc{T}_{\eta^{(0)}}$ 
connecting $\eta^{(0)}$ to $\muno$ and such that 
$H(\omega_i,\omega_i+1)<H(\muno)+\beta\Gamma$  
for any $i=0,\dots,n-1$. Finally, 
we remark that the path 
$\{\omega'_0=\omega_n,\omega'_1=\omega_{n-1}, 
         \dots,\omega'_n=\omega_0,\omega'_{n+1}=\eta\}$ 
satisfies the properties of point $i)$. 
A similar construction can be repeated for point $ii)$. 

Now, we come 
to the main points $iii)$ and $iv)$: our goal is to prove that  
$\Phi(\cc{G}_{\muno})=H(\muno)+\beta\Gamma$. First of all we 
notice that point $i)$ above implies 
\begin{displaymath} 
\Phi(\cc{G}_{\muno})\le\beta\Gamma+H(\muno)
\;\;\; , 
\end{displaymath} 
hence our calculation is reduced to prove the lower bound 
$\Phi(\cc{G}_{\muno})\ge\beta\Gamma+H(\muno)$.  
We have to examine all the paths connecting 
$\cc{G}_{\muno}$ with $\cc{S}\setminus \cc{G}_{\muno}$: such a path 
$\{\omega_0,\omega_1,\dots,\omega_n\}$ has at least a direct 
jump from $\cc{G}_{\muno}$ to $\cc{S}\setminus \cc{G}_{\muno}$, that is there 
exists $k\in\{0,\dots,n-1\}$ such that $\omega_k\in \cc{G}_{\muno}$ and 
$\omega_{k+1}\in\cc{S}\setminus\cc{G}_{\muno}$. Hence, for any $\omega$ 
connecting $\cc{G}_{\muno}$ with its exterior we have 
\begin{displaymath} 
\Phi_{\omega}\ge\min_{\sigma\in \cc{G}_{\muno},\eta\in\cc{S}\setminus 
\cc{G}_{\muno}} H(\sigma,\eta) 
\;\;\; . 
\end{displaymath} 

Given $\sigma\in\cc{G}_{\muno}$, we consider the configuration 
$\eta\in\cc{S}\setminus\cc{G}_{\muno}$ that can be reached with the 
smallest energetic cost, namely 
$\eta\in\cc{S}\setminus\cc{G}_{\muno}$ is such that 
\begin{equation}
\label{min}
H(\sigma,\eta)=\min_{\zeta\in\cc{S}\setminus\cc{G}_{\muno}}H(\sigma,\zeta)
\end{equation}
We have to prove that 
$H(\sigma,\eta)\ge\beta\Gamma+H(\muno)$ with the equality valid if and only if 
$\sigma\in\cc{O}_{\muno}$ and $\eta\in\pi_{\muno}(\sigma)$.  

First of all we note that 
$P_{\Lambda}(\sigma,\eta)
 \stackrel{\beta\to\infty}{\longrightarrow} 0$ 
otherwise we would have $\eta\in\cc{B}(\widehat{\sigma})$. From Table 1 we get 
that there exists $x\in\Lambda$ such that 
\begin{equation}
\label{st0}
\log p_x(\eta(x)|\sigma)=-2\beta(2-h)+o(e^{-\beta c})
\end{equation}
for some positive 
constant $c$. We consider, then, $\zeta=\eta^x$, and we remark that 
(\ref{min}) implies $\zeta\in\cc{G}_{\muno}$. 
Note that 
\begin{equation}
\label{st0.5}
\log P_{\Lambda}(\sigma,\eta)=
\log P_{\Lambda}(\sigma,\zeta)
-\log p_x(\zeta(x)|\sigma)
+\log p_x(\eta(x)|\sigma)
\end{equation}
that, together with (\ref{nhe}), implies
\begin{eqnarray}
\label{st0.7}
H(\sigma,\eta)
&=& H(\sigma,\zeta)-\log P_{\Lambda}(\sigma,\eta)
                   +\log P_{\Lambda}(\sigma,\zeta)
\nonumber\\
&=& H(\sigma,\zeta)-\log p_x(\eta(x)|\sigma)
                   +\log p_x(\zeta(x)|\sigma)
\nonumber\\
&\ge& H(\zeta)-\log p_x(\eta(x)|\sigma)
                   +\log p_x(\zeta(x)|\sigma)
\end{eqnarray}

We can characterize $\widehat\zeta$ as follows: by using 
Propositions \ref{p:res} and \ref{p:congl} we have that 
there exist $k\ge 1$ rectangles $R_{\ell_1,m_1},\dots,R_{\ell_k,m_k}$ 
satisfying the conditions of point $(iii)$ in Proposition \ref{p:res},
with respect to the configuration $\widehat\zeta$,  
and such that $\ell_i<\lcrit$ for any $i=1,\dots,k$. 

Let us consider $\zeta'\in\cc{S}$ such that 
$\zeta'_{\cup_{i=1}^k\core{R}_{\ell_i,m_i}}= 
 \muno_{\cup_{i=1}^k\core{R}_{\ell_i,m_i}}$ 
and
$\zeta'_{\Lambda\setminus\cup_{i=1}^k\core{R}_{\ell_i,m_i}}= 
 \zeta_{\Lambda\setminus\cup_{i=1}^k\core{R}_{\ell_i,m_i}}$. 
By recalling that there exist an uphill path joining $\widehat\zeta$ to
$\zeta$ it is easy to prove that there exist $s\ge 0$ 
rectangles
$R_{1,m_{k+1}},\dots,R_{1,m_{k+s}}$, with $m_{k+i}\ge 1$ 
for all $i=1,\dots,s$, such that 
$\zeta'_{\Lambda\setminus\cup_{i=1}^s\core{R}_{1,m_{k+i}}}= 
 \muno_{\Lambda\setminus\cup_{i=1}^s\core{R}_{1,m_{k+i}}}$ and 
$\zeta'$ coincides with a chessboard or $\puno$ inside 
$\core{R}_{1,m_{k+i}}$ for all $i=1,\dots,s$. 
 
We let $b_i^o$ (resp. $b_i^v$) the horizontal (resp. vertical) side length of
the rectangle $R_{\ell_i,m_i}$ for any $i=1,\dots,k+s$.
We set $b^o:=\sum_{i=1}^{k+s} b_i^o$, $b^v:=\sum_{i=1}^{k+s} b_i^v$ 
and we remark that 
\begin{equation}
\label{sper}
\sum_{i=1}^{k+s}(\ell_i+m_i)=b^o+b^v
\;\;\; .
\end{equation}

We suppose, now, $b^o+b^v\ge 2\lcrit$. 
By a direct evaluation of the energy of the rectangles
it is easy to show the bound 
\begin{eqnarray} 
\label{st1}
E(\zeta)-E(\muno)
&=& \left[E(\widehat\zeta)-E(\muno)\right]
   +\left[E(\zeta)-E(\widehat\zeta)\right]
\nonumber\\
&\ge& \left[E(\widehat\zeta)-E(\muno)\right]
     +\left[E(\zeta')-E(\muno)\right]
      \ge 4(b^o+b^v)-2h\sum_{i=1}^{k+s} b_i^o b_i^v 
\end{eqnarray}
Where we have used that two of the $R_{1,m_i}$, with $i=k+1,\dots,k+s$, 
rectangles can interact iff they are filled with different parity 
chessboards.
Now, by using the subcriticality of $\widehat\zeta$, 
namely by using $\ell_i<\lcrit$ for any $i=1,\dots,k$ (recall $\ell_i=1$ 
for all $i=k+1,\dots,k+s$), we have 
\begin{equation}
\label{st2}
\sum_{i=1}^{k+s} b_i^o b_i^v\le (\lcrit-1)[b^o+b^v-(\lcrit-1)]
\end{equation}
indeed, let $\ell=\max_{i=1,\dots,k}\ell_i$, we have
\begin{displaymath}
\sum_{i=1}^{k+s} b_i^o b_i^v
=
\sum_{i=1}^{k+s} \ell_i m_i
\le 
\ell\sum_{i=1}^{k+s} m_i 
\le
\ell [b^o+b^v-\ell]
\le
(\lcrit-1)[b^o+b^v-(\lcrit-1)]
\end{displaymath}
where, in the last inequality, we have used $b^o+b^v\ge 2\lcrit$. 
Now, 
recall $\lcrit=[2/h]+1=2/h+\varepsilon$ for some $\varepsilon\in(0,1)$,
then the inequality
\begin{equation}
\label{st0.9}
2-h(\lcrit-1)=2-h\left(\frac{2}{h}+\varepsilon-1\right)=h(1-\varepsilon)>0
\;\;\; ,
\end{equation}
(\ref{st1}), (\ref{st2}) and	
the hypothesis $b^o+b^v\ge 2\lcrit$ 
imply 
\begin{eqnarray}
\label{st3}
E(\zeta)-E(\muno) 
&\ge&   4(b^o+b^v)-2h(\lcrit-1)[b^o+b^v-(\lcrit-1)]
\nonumber\\
&=&     (b^o+b^v)[4-2h(\lcrit-1)]+2h(\lcrit-1)^2
\nonumber\\
&\ge&   2\lcrit[4-2h(\lcrit-1)]+2h(\lcrit-1)^2
\nonumber\\
&=&     8\lcrit-2h\lcrit^2+2h
\end{eqnarray}

Now, by using (\ref{st0}), (\ref{st0.7}) and (\ref{st3}) we get that 
for any $\delta>0$ there exist $\beta$ large enough such that 
\begin{equation}
\label{st4}
H(\sigma,\eta)-H(\muno)\ge  
8\beta\lcrit-2\beta h\lcrit^2+4\beta-\delta
\end{equation}
Finally, from the equation above it follows, by choosing $\delta$ small 
enough, that $H(\sigma,\eta)>\beta\Gamma+H(\muno)$.

We come, now, to the case $b^o+b^v\le 2\lcrit-1$. 
First of all we notice that $\zeta$ and $\eta$ differ for the value of
a single spin, this implies that in $\eta$ there is a single supercritical 
rectangle. More precisely,
there exist $k\ge 1$ rectangles 
$R_{\ell'_1,m'_1},\dots,R_{\ell'_{k'},m'_{k'}}$ 
satisfying the conditions of point $(iii)$ in Proposition \ref{p:res},
with respect to the configuration $\widehat\eta$,  
and such that $\ell'_1\ge\lcrit$ and $\ell'_i<\lcrit$ for any $i=2,\dots,k'$. 

Let $\ell^o$ (resp.\ $\ell^v$) the length of the horizontal (resp.\ vertical) 
side of the supercritical rectangle $R_{\ell'_1,m'_1}$. We note that 
$b^o+b^v\le 2\lcrit-1$ implies that $\ell^o+\ell^v$ is surely less than 
$4\lcrit$; one can show, indeed, that $\ell^o+\ell^v$ does not 
exceed $2\lambda+4$. Under this condition it is easy to show the bound 
\begin{equation}
\label{st5}
E(\widehat\eta)-E(\muno)\ge 4(\ell^o+\ell^v)-2h\ell^o\ell^v
\;\;\; ,
\end{equation}
indeed the energy of $\widehat\eta$ can be bounded from below with the 
energy of an $\ell^o\times\ell^v$ chessboard droplet in the sea of minuses. 

Now, let $C\in\{\eC,\oC\}$ such that $\eta(x)=C(x)$ and consider 
the collection of rectangles 
$\cc{R}\subset\{R_{\ell_1,m_1},\dots,R_{\ell_k,m_k},R_{1,m_{k+1}},
                \dots,R_{1,m_{k+s}}\}$,
such that for each $R\in\cc{R}$ either 
$\widehat\zeta_{\core{R}}=\puno_{\core{R}}$ or the chessboard part of 
$\core{R}$ coincides with $C$. By remarking that $\cc{R}$ is a collection of 
pairwise not interacting rectangles, 
we can find a positive integer $\Delta\le 9$ such that 
if we let 
\begin{equation}
\label{st6}
V:=
\core{R}_{\ell'_1,m'_1}\setminus
\bigcup_{R\in\cc{R}}\core{R}
\;\;\;\;\;\textrm{and}\;\;\;\;\;
N:=\left|
\core{R}_{\ell'_1,m'_1}\setminus
\bigcup_{R\in\cc{R}}\core{R}\right|=|V|
\end{equation}
we get the lower bound 
\begin{equation}
\label{st7}
E(\eta)-E(\muno)\ge \left[E(\widehat\eta)-E(\muno)\right]+2h(N-\Delta)
                \ge 4(\ell^o+\ell^v)-2h\ell^o\ell^v+2h(N-\Delta)
\end{equation}
where, in the last inequality, we have used (\ref{st5}).
We remark that $N$ is a lower bound of 
the number of sites in $R_{\ell'_1,m'_1}$ not 
belonging to any cluster of $\zeta$ that will persist in $\widehat\eta$. 

We consider, now, the geometrical projection of the rectangles 
$R_{\ell_i,m_i}$, with $i=1,\dots,k+s$, onto one of the horizontal (resp.\
vertical) sides of $R_{\ell'_1,m'_1}$. Such a projection is a collection of,
maybe not disjoint, segments; we denote with $p^o$ (resp. $p^v$) the 
length of the union of these segments. By definition we have $p^o\le b^o$ 
(resp.\ $p^v\le b^v$). We mention the following interesting bound on $N$:
\begin{equation}
\label{st8}
N\ge\ell^o\ell^v-p^o p^v
\end{equation}
Indeed, $\ell^v$ sites of $V$ are associated to each unit segment of 
the horizontal side of $R_{\ell'_1,m'_1}$ not belonging to the 
the projection of the rectangles $R_{\ell_i,m_i}$. Moreover, 
$p^o$ (not already counted) 
sites of $V$ are associated to each unit segment of 
the vertical side of $R_{\ell'_1,m'_1}$ not belonging to the 
the projection of the rectangles $R_{\ell_i,m_i}$.
Hence,
\begin{displaymath}
N=|V|\ge (\ell^o-p^o)\ell^v+(\ell^v-p^v)p^o=\ell^o\ell^v-p^op^v 
\;\;\; .
\end{displaymath}

Without loss of generality we can, now, suppose $b^o\le b^v$. This implies
$b^o\le\lcrit-1$, indeed if it were, by absurdity, $b^o\ge\lcrit$, then
it would be $b^o+b^v\ge 2\lcrit>2\lcrit-1$. We distinguish
among four different situations.  

\smallskip\par\noindent
\textit{Case 1.} $p^o\le\lcrit-2$. By inserting (\ref{st8}) in 
(\ref{st7}) we get 
\begin{eqnarray}
\label{st9}
E(\eta)-E(\muno)
&\ge& 4(\ell^o+\ell^v)-2h\ell^o\ell^v+2h(\ell^o\ell^v-p^op^v)-2h\Delta
\nonumber\\
&\ge& 4(\ell^o+\ell^v)-2h(\lcrit-2)\ell^v-2h\Delta
\end{eqnarray}
where, in the last inequality, we use $p^o\le\lcrit-2$ and $p^v\le\ell^v$. 
Now, recalling $\ell^v\ge\lcrit$, for $h$ small enough we have 
$\ell^v\ge\Delta$, hence 
\begin{eqnarray}
\label{st10}
E(\eta)-E(\muno)
&\ge& 4(\ell^o+\ell^v)-2h(\lcrit-1)\ell^v=4\ell^o+2\ell^v[2-h(\lcrit-1)]
\nonumber\\
&\ge& 4\lcrit+2\lcrit[2-h(\lcrit-1)]=8\lcrit-2h\lcrit^2+2h\lcrit>\Gamma
\end{eqnarray}
where we have used (\ref{st0.9}).

\smallskip\par\noindent
\textit{Case 2.} $p^o=\lcrit-1$ (recall $p^o\le\lcrit-1$) and 
$(\ell^o,\ell^v)\not=(\lcrit,\lcrit)$. As in Case 1 we get
\begin{equation}
\label{st11}
E(\eta)-E(\muno)
\ge 4(\ell^o+\ell^v)-2hp^op^v-2h\Delta
\ge 4(\ell^o+\ell^v)-2h(\lcrit-1)\lcrit-2h\Delta
\end{equation}
where in the last inequality we have used $p^o=\lcrit-1$ and 
$p^v\le\lcrit$, indeed 
\begin{displaymath}
p^v\le b^v\le 2\lcrit-1-b^o\le 2\lcrit-1-p^o=\lcrit
\end{displaymath}
Hence, by using $\ell^o+\ell^v\ge 2\lcrit+1$ we get
\begin{eqnarray}
\label{st12}
E(\eta)-E(\muno)
&\ge& 4(2\lcrit+1)-2h(\lcrit-1)\lcrit-2h\Delta
\nonumber\\
&=&   8\lcrit-2h\lcrit^2+2h(\lcrit-1)+2h(\lcrit-\Delta)+[4-2h(\lcrit-1)]
\nonumber\\
&=&   \Gamma+2h(\lcrit-\Delta)+[4-2h(\lcrit-1)]>\Gamma
\end{eqnarray}
where we have used $4-2h(\lcrit-1)>0$ (see inequality (\ref{st0.9})
and have chosen
$h$ small enough in order to get $\lcrit>\Delta$. 

\smallskip\par\noindent
\textit{Case 3.} $p^o=\lcrit-1$, $\ell^o=\ell^v=\lcrit$ and $k+s\ge 2$. 
In this case we have the easy estimate (see \cite{[KO]} and \cite{[NO]}) 
$N\ge 2\lcrit$,
hence from (\ref{st7}) we get 
\begin{equation}
\label{st13}
E(\eta)-E(\muno)
\ge 8\lcrit-2h\lcrit^2+2h(2\lcrit-\Delta)
\;\;\; .
\end{equation}
By choosing $h$ small enough we get $E(\eta)>\Gamma+E(\muno)$. 
\smallskip\par\noindent
\textit{Case 4.} $p^o=\lcrit-1$, $\ell^o=\ell^v=\lcrit$ and $k+s=1$.
We have to consider the three configurations 
$\eta_1$, $\eta_2$ and $\eta_3$ 
depicted in Fig.~\ref{f:fin}, 
where
the $\lcrit\times(\lcrit-1)$ rectangle is filled with one of the two
chessboards and $\eta(x)=+1$. By a 
direct evaluation of the energy we have that 
\begin{displaymath}
E(\eta_2)=\Gamma-4h,
\;\;\;\;
E(\eta_1)-E(\eta_2)=4+2h>4h
\;\;\;\;\textrm{ and }\;\;\;\;
E(\eta_3)-E(\eta_2)=2h(\lcrit-1)>4h
\;\;\; .
\end{displaymath}
Now, starting from $\eta_2$ the lowest energy jump toward $\cc{G}_{\muno}$ 
consists in reverting all the spins inside the rectangle $R$ and the plus 
spin associated to $x$. The cost of such a jump is $\exp{(-4\beta h)}$, indeed 
we have to pay in order to keep the two minuses around $x$. Finally we have
that $H(\sigma,\eta)=\beta\Gamma+H(\muno)$ iff 
$\sigma\in\cc{O}_{\muno}$ and $\eta\in\pi_{\muno}(\sigma)$.
\qed 

\setlength{\unitlength}{1.3pt} 
\begin{figure} 
\begin{picture}(200,200)(-80,-50) 
\thinlines 
\put(-60,50){\line(0,1){80}} 
\put(-60,50){\line(1,0){80}} 
\put(-60,130){\line(1,0){80}} 
\put(20,50){\line(0,1){80}} 
\qbezier[40](10,50)(10,90)(10,130)
\put(-25,30){$\eta_1$} 
\put(-25,135){$\lcrit$} 
\put(-70,85){$\lcrit$} 
\put(23,88){$x$} 
\put(20,85){\line(0,1){10}} 
\put(20,85){\line(1,0){10}} 
\put(20,95){\line(1,0){10}} 
\put(30,85){\line(0,1){10}} 
\put(60,50){\line(0,1){80}} 
\put(60,50){\line(1,0){80}} 
\put(60,130){\line(1,0){80}} 
\put(140,50){\line(0,1){80}} 
\qbezier[40](130,50)(130,90)(130,130)
\put(95,30){$\eta_2$} 
\put(95,135){$\lcrit$} 
\put(50,85){$\lcrit$} 
\put(133,88){$x$} 
\put(130,85){\line(0,1){10}} 
\put(130,85){\line(1,0){10}} 
\put(130,95){\line(1,0){10}} 
\put(140,85){\line(0,1){10}} 
\put(180,50){\line(0,1){80}} 
\put(180,50){\line(1,0){80}} 
\put(180,130){\line(1,0){80}} 
\put(260,50){\line(0,1){80}} 
\qbezier[40](250,50)(250,90)(250,120)
\qbezier[40](250,120)(210,120)(180,120)
\put(215,30){$\eta_3$} 
\put(215,135){$\lcrit$} 
\put(170,85){$\lcrit$} 
\put(253,123){$x$} 
\put(250,120){\line(0,1){10}} 
\put(250,120){\line(1,0){10}} 
\put(250,130){\line(1,0){10}} 
\put(260,120){\line(0,1){10}} 
\end{picture} 
\vskip -3 cm 
\caption[minmax]{The three possible situations
that must be taken into account in Case 4.}
\label{f:fin} 
\end{figure}
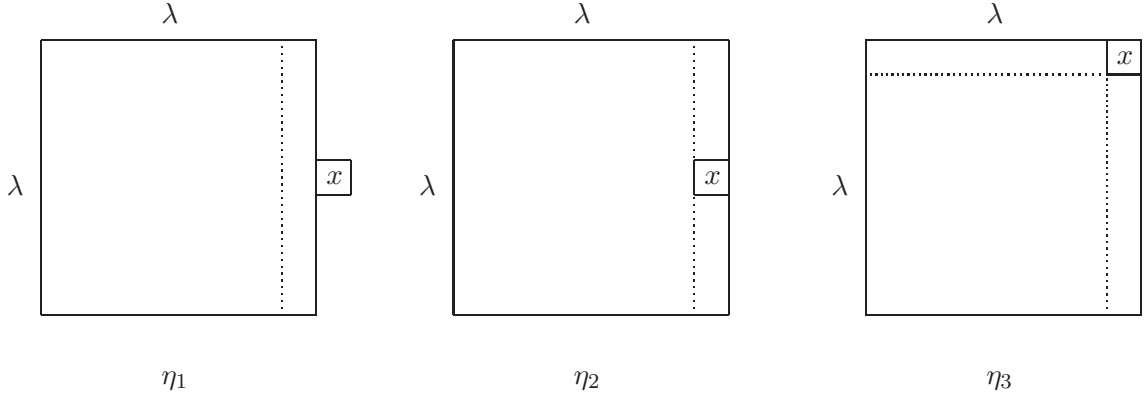 
\sezione{Proof of the Propositions}{s:dimp} 
\par\noindent 
In this section we prove the Propositions stated throughout the paper: 
the tools 
which will be used are those outlined in Subsection \ref{su:strum}. 
\par\noindent 
The Proposition \ref{p:std} is a straightforward consequence 
of Lemma \ref{l:sta} and the characterization of the local minima 
of the energy given in \cite{[NS1]}. 
 
\addcontentsline{toc}{subsection}{{\it Proof of Proposition \ref{p:res}}} 
\medskip\noindent{\it Proof of Proposition \ref{p:res}.}\/ 
We just give a sketch of the proof.
$i)$ Let $\sigma\in\cc{S}_{\puno}\setminus\{\puno\}$. 
There exists $x\in\Lambda$ with two neighboring sites in the sea of pluses
such that $\sigma(x)=-1$, then we will have $\rmT^n\sigma(x)=+1$ for all
$n\ge 1$, hence $\sigma$ is not an element of a stable pair.
$ii)$ Let $C\in\{\eC,\oC\}$ and $\sigma\in\cc{S}_C$. Suppose all the sites
of $\Lambda$ not belonging to the sea of chessboard are occupied by pluses,
and suppose these pluses form a single
cluster $X\subset\Lambda$. Consider the maximal $Y\subset X$ such that 
for each $y\in Y$ there exists a $2\times 2$ subset of $Y$ containing $y$.
If $Y$ is not rectangular shaped, then there exists $x$ such that 
$\sigma(x)=-1$ and at least two among its neighbors belong to $Y$. Then 
$\rmT^n\sigma(x)=+1$ for all $n\ge 1$ implies $\sigma$ is not an element of a 
stable pair. 
The proof can be easily generalized to the case with more than a 
cluster of pluses.
$iii)$ The proof is similar to the one sketched for the case $ii)$. 
\qed\smallskip\par 
 
\addcontentsline{toc}{subsection}{{\it Proof of Proposition 
\ref{p:contr}}} 
\medskip\noindent{\it Proof of Proposition \ref{p:contr}.}\/ 
Let us consider a rectangle $R_{\ell,m}$ with $2\le\ell\le m\le L-2$. 
\smallskip\par\noindent 
{\it Case 1}: 
let $\eta\in\cc{S}_{\muno}$ be the trap such that 
$\eta_{\Lambda\setminus\core{R}_{\ell,m}}= 
\muno_{\Lambda\setminus\core{R}_{\ell,m}}$ and 
$\eta_{\core{R}_{\ell,m}}=\oC_{\core{R}_{\ell,m}}$; suppose $\ell<\lcrit$. 
Idea of the proof: 
we characterize the basin $\cc{B}(\eta)$, that is we find 
$\Upsilon(\eta)$; 
then we suppose the system prepared in 
$\eta$ (namely  $\sigma_0=\eta$) and, by means of Lemmata \ref{l:enzo0} and 
\ref{l:enzo}, we estimate both 
$\tau_{\cc{S}\setminus\overline{\cc{B}}(\sigma)}$ and 
$\sigma_{\tau_{\cc{S}\setminus\overline{\cc{B}}(\sigma)}}$. 

By using (\ref{caut}), 
$\Upsilon(\eta)$ can be estimated via the construction 
of the paths in $\Xi(\eta)$. 
First of all we consider all the possible transitions that can be 
first steps for a path in $\Xi(\eta)$: 
\begin{enumerate} 
\item 
the configuration on $\core{R}_{\ell,m}$ is flipped together with a minus spin in 
$\Lambda\setminus\core{R}_{\ell,m}$ adjacent to one of the plus spins in 
$\core{R}_{\ell,m}$. 
A configuration $\eta_1\in\cc{S}\setminus \cc{B}(\eta)$ 
is reached and 
$H(\eta,\eta_1)-H(\eta)=2\beta(2-h)=:\Phi_1$. Hence 
\begin{equation} 
\Upsilon(\eta)\le 
H(\eta)+\Phi_1= 
H(\eta)+2\beta(2-h) 
\;\;\; . 
\label{phi1} 
\end{equation} 
As shown in Fig. \ref{f:cre} the unique downhill path starting from 
$\eta_1$ ends in the trap $\widehat\eta_1$ coincident with a 
chessboard inside a rectangle $R_{\ell+1,m}$ 
(or $R_{\ell,m+1}$, depending on which side the protuberance appeared) and 
with $\muno$ outside. 
We notice that the energy of $\eta_1$ depends whether the plus 
protuberance appears in the middle of one side or on the corner, but 
$H(\eta,\eta_1)$ does not depend on this detail. 
\item 
The configuration on $\core{R}_{\ell,m}$ is flipped together with a minus spin in 
$\Lambda\setminus\core{R}_{\ell,m}$ 
with four neighboring pluses. 
A configuration $\eta_2\in \cc{B}(\eta)$ is reached such 
that 
$H(\eta,\eta_2)-H(\eta)=2\beta(4-h)>\Phi_1$: this kind of 
steps can be neglected. 
\item 
All the spins inside $\core{R}_{\ell,m}$ are flipped excepted one corner minus 
(if all the corner are pluses, then this step is considered after a full 
flip of the configuration inside $\core{R}_{\ell,m}$). 
A configuration $\eta_3\in \cc{B}(\eta)$ is reached such that 
$H(\eta,\eta_3)-H(\eta)=2\beta h$. 
\item 
All the spins inside $\core{R}_{\ell,m}$ are flipped excepted one minus in the 
middle (not on the corner) of one of the four sides of the rectangle 
(if such a spin does not exist, this can happen in the case $\ell=m=3$, 
then this step is considered after a full 
flip of the configuration inside $\core{R}_{\ell,m}$). 
A configuration $\eta_4\in \cc{B}(\eta)$ is reached such that 
$H(\eta,\eta_4)-H(\eta)=2\beta(2+h)>\Phi_1$: 
this kind of steps can be neglected. 
\item 
All the spins inside $\core{R}_{\ell,m}$ are flipped excepted one minus 
with four nearest neighboring pluses 
(if such a spin does not exist, this can happen in the case $\ell=m=3$, 
then this step is considered after a full 
flip of the configuration inside $\core{R}_{\ell,m}$). 
A configuration $\eta_5\in \cc{B}(\eta)$ is reached such that 
$H(\eta,\eta_5)-H(\eta)=2\beta(4+h)>\Phi_1$: 
this kind of steps can be neglected. 
\item 
All the spins inside $\core{R}_{\ell,m}$ are flipped excepted one plus spin. 
A configuration $\eta_6\in \cc{B}(\eta)$ is reached such that 
$H(\eta,\eta_6)-H(\eta)=2\beta(4-h)>\Phi_1$: 
this kind of steps can be neglected. 
\item 
Two or more events among those listed above are performed simultaneously: 
the energy cost 
is smaller than $\Phi_1$ only in the case of a simultaneous 
persistence of $k$ minus corners of the chessboard. 
All the others multiple events can be neglected. 
\end{enumerate} 
From the list above it follows that there exists a path $\omega'= 
\{\eta,\eta_1\}\in\Xi(\eta)$ consisting of a single step of the first 
type; so $\Phi_{\omega'}=H(\eta)+\Phi_1=H(\eta)+2\beta(2-h)$. The only paths 
$\omega\in\Xi(\eta)$ that can compete with $\omega'$ are those 
whose first step is a single or a multiple minus corner persistence. 
After such a step (see, for instance, the first step in 
Fig. \ref{f:con}) the configuration 
is a chessboard on a 
subset of $\core{R}_{\ell,m}$ obtained by removing some of the four corners 
of $\core{R}_{\ell,m}$ and $\muno$ outside. 
By a direct inspection it follows that starting from this configuration 
the possible second steps of our paths are exactly those listed above: no 
new step enters into the game. 

By iterating the above argument, it follows that a path 
$\omega''\in\Xi(\eta)$ 
such that $\Phi_{\omega''}\le\Phi_{\omega'}$ 
can be obtained 
by using only single or multiple minus corner persistences. 
The best path $\omega''$ 
is a sequence of $\ell-1$ minus corner persistences 
on one of the two sides of the rectangle long $\ell$: 
$\Phi_{\omega''}=H(\eta)+2\beta h(\ell-1)$. 
By comparing $\Phi_{\omega'}$ and 
$\Phi_{\omega''}$, and recalling that $\ell<\lcrit$, one 
obtains 
$\Upsilon(\eta)=H(\eta)+2\beta h(\ell-1)$. By Lemma \ref{l:enzo0} we 
obtain 
$\Phi(\overline{\cc{B}}(\eta))=H(\eta)+2\beta h(\ell-1)$ and 
$U(\overline{\cc{B}}(\eta))=\{\eta''\}$, 
where $\eta''$ is a configuration coincident 
with a chessboard in a rectangle $R_{\ell,m-1}$ and with $\muno$ 
outside. 
Finally, by applying Lemma \ref{l:enzo} we can estimate 
$\tau_{\cc{S}\setminus\overline{\cc{B}}(\eta)}\sim\exp\{2\beta h(\ell-1)\}$ 
and we obtain that with 
high probability 
$\sigma_{\tau_{\cc{S}\setminus\overline{\cc{B}}(\eta)}}=\eta''$. 
By using the Markov property and by 
iterating the argument above one completes the proof of part $i)$ of 
Proposition \ref{p:contr}. The proof of part $ii)$ is similar. 
\smallskip\par\noindent 
{\it Case 2}: 
Let us consider the trap $\eta\in\cc{S}_{\muno}$ such that 
$\eta_{\Lambda\setminus\core{R}_{\ell,m}}= 
\muno_{\Lambda\setminus\core{R}_{\ell,m}}$ and 
$\eta_{\core{R}_{\ell,m}}=\eC_{\core{R}_{\ell,m}}$. 
The proof is the same as in the Case 1. 
\smallskip\par\noindent 
{\it Case 3}: 
Let us consider the trap $\eta\in\cc{S}_{\muno}$ such that 
$\eta_{\Lambda\setminus\core{R}_{\ell,m}}= 
\muno_{\Lambda\setminus\core{R}_{\ell,m}}$ and 
$\eta_{\core{R}_{\ell,m}}=\puno_{\core{R}_{\ell,m}}$. Again we 
suppose $\ell<\lcrit$. 
As before we start by listing the transitions that can be 
first steps for a path in $\Xi(\eta)$: 
\newcounter{lista} 
\begin{list} 
{\arabic{lista}.}{ 
\usecounter{lista} 
\setlength{\labelwidth}{2cm} 
} 
\item 
a minus spin adjacent to one of the four sides of the rectangle 
is flipped. A configuration $\eta_1\in\cc{S}\setminus \cc{B}(\eta)$ 
is reached and 
$H(\eta,\eta_1)-H(\eta)=2\beta(2-h)=:\Phi_1$. 
We notice that the unique downhill path starting from 
$\eta_1$ ends in a trap $\eta'$ as in Fig. \ref{f:crep}. 
\setlength{\unitlength}{1pt} 
\begin{figure} 
\begin{picture}(200,150)(-250,-40) 
\thinlines 
\put(-130,90){$+$}\put(-120,90){$+$} 
\put(-130,80){$+$}\put(-120,80){$+$} 
\put(-130,70){$+$}\put(-120,70){$+$} 
\put(-130,60){$+$}\put(-120,60){$+$} 
\put(-130,50){$+$}\put(-120,50){$+$} 
\put(-130,40){$+$}\put(-120,40){$+$} 
\put(-135,37){\line(1,0){25}} 
\put(-110,37){\line(0,1){63}} 
\put(-135,100){\line(1,0){25}} 
\put(-100,60){$\stackrel{\scriptstyle e^{-2\beta(2-h)}}{\longrightarrow}$} 
\put(-60,90){$+$}\put(-50,90){$+$} 
\put(-60,80){$+$}\put(-50,80){$+$}\put(-40,80){$-$} 
\put(-60,70){$+$}\put(-50,70){$+$}\put(-40,70){$+$} 
\put(-60,60){$+$}\put(-50,60){$+$}\put(-40,60){$-$} 
\put(-60,50){$+$}\put(-50,50){$+$} 
\put(-60,40){$+$}\put(-50,40){$+$} 
\put(-65,37){\line(1,0){25}} 
\put(-40,37){\line(0,1){20}} 
\put(-40,57){\line(1,0){10}} 
\put(-30,57){\line(0,1){32}} 
\put(-30,89){\line(-1,0){10}} 
\put(-40,89){\line(0,1){11}} 
\put(-65,100){\line(1,0){25}} 
\put(-25,60){$\stackrel{\scriptstyle\phantom{e^{-2\beta h}}}{\longrightarrow}$} 
\put(5,90){$+$}\put(15,90){$+$}\put(25,90){$-$} 
\put(5,80){$+$}\put(15,80){$+$}\put(25,80){$+$} 
\put(5,70){$+$}\put(15,70){$+$}\put(25,70){$-$} 
\put(5,60){$+$}\put(15,60){$+$}\put(25,60){$+$} 
\put(5,50){$+$}\put(15,50){$+$}\put(25,50){$-$} 
\put(5,40){$+$}\put(15,40){$+$} 
\put(0,37){\line(1,0){25}} 
\put(25,37){\line(0,1){10}} 
\put(25,47){\line(1,0){10}} 
\put(35,47){\line(0,1){53}} 
\put(0,100){\line(1,0){35}} 
\put(40,60){$\stackrel{\scriptstyle\phantom{e^{-2\beta h}}}{\longrightarrow}$} 
\put(70,90){$+$}\put(80,90){$+$}\put(90,90){$+$} 
\put(70,80){$+$}\put(80,80){$+$}\put(90,80){$-$} 
\put(70,70){$+$}\put(80,70){$+$}\put(90,70){$+$} 
\put(70,60){$+$}\put(80,60){$+$}\put(90,60){$-$} 
\put(70,50){$+$}\put(80,50){$+$}\put(90,50){$+$} 
\put(70,40){$+$}\put(80,40){$+$}\put(90,40){$-$} 
\put(65,37){\line(1,0){35}} 
\put(100,37){\line(0,1){63}} 
\put(65,100){\line(1,0){35}} 
\end{picture} 
\vskip -2.5 cm  
\caption[Pluses droplet growth]{Growth of a plus droplet inside 
the sea of minuses: appearing of a protuberance.} 
\label{f:crep} 
\end{figure}
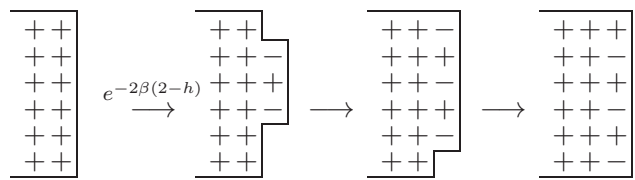 
\item 
A minus spin at distance greater or equal to two from any site of the 
rectangle is flipped. A configuration $\eta_2\in \cc{B}(\eta)$ is reached such 
that 
$H(\eta,\eta_2)-H(\eta)=2\beta(4-h)>\Phi_1$: 
this kind of steps can be neglected. 
\item 
One of the four corners of the rectangle is flipped (a corner is a 
plus spin with two minuses among its nearest neighbors). 
A configuration $\eta_3\in \cc{B}(\eta)$ is reached such that 
$H(\eta,\eta_3)-H(\eta)=2\beta h$. 
\item 
One of the non--corner plus spin on one of the sides of the 
rectangle is flipped. 
A configuration $\eta_4\in \cc{B}(\eta)$ is reached such 
that 
$H(\eta,\eta_4)-H(\eta)=2\beta(2+h)>\Phi_1$: 
this kind of steps can be neglected. 
\item 
One plus spin in the interior of the rectangle is flipped. 
A configuration $\eta_5\in \cc{B}(eta)$ is reached such 
that 
$H(\eta,\eta_5)-H(\eta)=2\beta(4+h)>\Phi_1$: 
this kind of steps can be neglected. 
\item 
Two or more spins are flipped simultaneously: the energy cost 
is smaller than $\Phi_1$ only in the case of a simultaneous 
flip of $k$ corners. All the others many--spin flips can be neglected. 
\end{list} 
From the list above it follows that there exists a path $\omega'= 
\{\eta,\eta_1\}\in\Xi(\eta)$ consisting of a single step of the first 
type; so $\Phi_{\omega'}=H(\eta)+\Phi_1=H(\eta)+2\beta(2-h)$. The only paths 
$\omega\in\Xi(\eta)$ that can compete with $\omega'$ are those 
whose first step is a single or a multiple corner erosion. 
Suppose that after the first step of our uphill path the configuration 
of the system is $\eta_3$. 
Two more possible transitions must be taken into account 
in the analysis of the possible second steps: 
\newcounter{lista1} 
\begin{list} 
{\arabic{lista1}.}{ 
\usecounter{lista1} 
\setlength{\labelwidth}{2cm} 
} 
\setcounter{lista1}{\thelista} 
\item 
one corner spin is flipped. A configuration $\eta_7\in \cc{B}(\eta)$ 
is reached such that 
$H(\eta_3,\eta_7)-H(\eta_3)=4\beta h$. Indeed we have 
to take into account that the minus spin with two pluses among its 
nearest neighbors (the minus at the site flipped at the first step) must 
persist. 
\item 
The minus spin with two pluses among its 
nearest neighbors and one of its two adjacent plus spins are 
simultaneously flipped. 
A configuration $\eta_8\in \cc{B}(\eta)$ 
is reached such that $H(\eta_3,\eta_8)-H(\eta_3)=2\beta h$. 
\end{list} 
From the third step on no more possible transitions arise, excepted the 
obvious generalization of 8: 
\newcounter{listab} 
\begin{list} 
{\arabic{listab}.}{ 
\usecounter{listab} 
\setlength{\labelwidth}{2cm} 
} 
\setcounter{listab}{8} 
\item 
a corner plus spin at site $x$ is flipped together with 
all the spins at sites $y\not= x$ such that 
$p_y(\eta^y(y)|\eta)\longrightarrow 1$ in the limit $\beta\to\infty$, where 
$\eta$ denotes the actual configuration. 
The energy cost of this transition is $2\beta h$. 
\end{list} 
We conclude that an 
estimate of $\Upsilon(\eta)$ smaller then 
$\Phi_{\omega'}=H(\eta)+2\beta(2-h)$ can be obtained only 
by using an uphill path made of steps of types 3, 7, 8 and 9, or steps 
in which two or more transitions 3, 7, 8, 9 are performed simultaneously. 
Consider a path obtained by using these transitions: 
until on each side of the rectangle there are two nearest neighboring pluses 
the configuration is still in $\cc{B}(\eta)$. Hence, to exit $\cc{B}(\eta)$ 
at least on one of the four sides of the rectangle there must be no 
pair of nearest neighboring plus spins. It is clear that the path $\omega''$ 
made 
of steps 3, 7, 8 and 9, exiting $\cc{B}(\eta)$ and with minimal 
height along the path is the one described in Fig. \ref{f:conp}: 
after a first 
step of type 3 and a second step of type 8, $l-3$ steps of type 9 
are performed until the stable pair $\eta''$ is reached. 
The height along this path is $\Phi_{\omega''}=H(\eta)+2\beta h(\ell-1)$. 

By comparing $\Phi_{\omega'}$ and $\Phi_{\omega''}$ 
and recalling that $\ell<\lcrit$, one 
obtains 
$\Upsilon(\eta)=\Phi_{\omega''}=H(\eta)+2\beta h(\ell-1)$; 
by Lemma \ref{l:enzo0} we obtain 
$\Phi(\overline{\cc{B}}(\eta))=H(\eta)+2\beta h(\ell-1)$ and 
$U(\overline{\cc{B}}(\eta))=\{\eta''\}$. 
Finally, by applying Lemma \ref{l:enzo} we can estimate 
$\tau_{\cc{S}\setminus\overline{\cc{B}}(\eta)}\sim\exp\{2\beta h(\ell-1)\}$ 
and we obtain that with 
high probability $\sigma_{\tau_{\cc{S}\setminus \cc{B}(\eta)}}=\eta''$. 
By using the Markov property and the results proven in the Case 1 
one completes the proof of part $i)$ of 
Proposition \ref{p:contr}. The proof of part $ii)$ is similar. 
\qed\smallskip\par 
\setlength{\unitlength}{1pt} 
\begin{figure} 
\begin{picture}(200,150)(-190,-40) 
\thinlines 
\put(-130,90){$+$}\put(-120,90){$+$}\put(-110,90){$+$} 
\put(-130,80){$+$}\put(-120,80){$+$}\put(-110,80){$+$} 
\put(-130,70){$+$}\put(-120,70){$+$}\put(-110,70){$+$} 
\put(-130,60){$+$}\put(-120,60){$+$}\put(-110,60){$+$} 
\put(-130,50){$+$}\put(-120,50){$+$}\put(-110,50){$+$} 
\put(-130,40){$+$}\put(-120,40){$+$}\put(-110,40){$+$} 
\put(-135,37){\line(1,0){35}} 
\put(-100,37){\line(0,1){63}} 
\put(-135,100){\line(1,0){35}} 
\put(-90,60){$\stackrel{\scriptstyle e^{-2\beta h}}{\longrightarrow}$} 
\put(-60,90){$+$}\put(-50,90){$+$}\put(-40,90){$-$} 
\put(-60,80){$+$}\put(-50,80){$+$}\put(-40,80){$+$} 
\put(-60,70){$+$}\put(-50,70){$+$}\put(-40,70){$+$} 
\put(-60,60){$+$}\put(-50,60){$+$}\put(-40,60){$+$} 
\put(-60,50){$+$}\put(-50,50){$+$}\put(-40,50){$+$} 
\put(-60,40){$+$}\put(-50,40){$+$}\put(-40,40){$+$} 
\put(-65,37){\line(1,0){35}} 
\put(-30,37){\line(0,1){63}} 
\put(-65,100){\line(1,0){35}} 
\put(-25,60){$\stackrel{\scriptstyle e^{-2\beta h}}{\longrightarrow}$} 
\put(5,90){$+$}\put(15,90){$+$}\put(25,90){$+$} 
\put(5,80){$+$}\put(15,80){$+$}\put(25,80){$-$} 
\put(5,70){$+$}\put(15,70){$+$}\put(25,70){$+$} 
\put(5,60){$+$}\put(15,60){$+$}\put(25,60){$+$} 
\put(5,50){$+$}\put(15,50){$+$}\put(25,50){$+$} 
\put(5,40){$+$}\put(15,40){$+$}\put(25,40){$+$} 
\put(0,37){\line(1,0){35}} 
\put(35,37){\line(0,1){63}} 
\put(0,100){\line(1,0){35}} 
\put(40,60){$\stackrel{\scriptstyle e^{-2\beta h}}{\longrightarrow}$} 
\put(70,90){$+$}\put(80,90){$+$}\put(90,90){$-$} 
\put(70,80){$+$}\put(80,80){$+$}\put(90,80){$+$} 
\put(70,70){$+$}\put(80,70){$+$}\put(90,70){$-$} 
\put(70,60){$+$}\put(80,60){$+$}\put(90,60){$+$} 
\put(70,50){$+$}\put(80,50){$+$}\put(90,50){$+$} 
\put(70,40){$+$}\put(80,40){$+$}\put(90,40){$+$} 
\put(65,37){\line(1,0){35}} 
\put(100,37){\line(0,1){63}} 
\put(65,100){\line(1,0){35}} 
\put(105,60){$\stackrel{\scriptstyle e^{-2\beta h}}{\longrightarrow}$} 
\put(135,90){$+$}\put(145,90){$+$}\put(155,90){$+$} 
\put(135,80){$+$}\put(145,80){$+$}\put(155,80){$-$} 
\put(135,70){$+$}\put(145,70){$+$}\put(155,70){$+$} 
\put(135,60){$+$}\put(145,60){$+$}\put(155,60){$-$} 
\put(135,50){$+$}\put(145,50){$+$}\put(155,50){$+$} 
\put(135,40){$+$}\put(145,40){$+$}\put(155,40){$+$} 
\put(130,37){\line(1,0){35}} 
\put(165,37){\line(0,1){63}} 
\put(130,100){\line(1,0){35}} 
\put(170,60){$\stackrel{\scriptstyle e^{-2\beta h}}{\longrightarrow}$} 
\put(200,90){$+$}\put(210,90){$+$}\put(220,90){$-$} 
\put(200,80){$+$}\put(210,80){$+$}\put(220,80){$+$} 
\put(200,70){$+$}\put(210,70){$+$}\put(220,70){$-$} 
\put(200,60){$+$}\put(210,60){$+$}\put(220,60){$+$} 
\put(200,50){$+$}\put(210,50){$+$}\put(220,50){$-$} 
\put(200,40){$+$}\put(210,40){$+$}\put(220,40){$+$} 
\put(195,37){\line(1,0){35}} 
\put(230,37){\line(0,1){63}} 
\put(195,100){\line(1,0){35}} 
\end{picture} 
\vskip -2.5 cm  
\caption[Pluses droplet contraction]{Shrinking of a droplet of pluses 
inside the sea of minuses.} 
\label{f:conp} 
\end{figure}
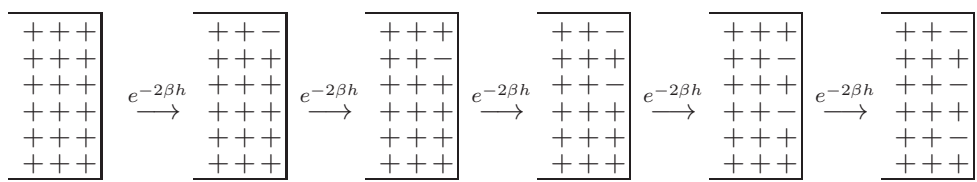 
\noindent 
The Propositions \ref{p:conch} and \ref{p:congl} can be proven 
via arguments similar to those used in the proof of Proposition 
\ref{p:contr}. Proposition \ref{p:cre} is a byproduct 
of the proves of Propositions \ref{p:contr} and 
\ref{p:conch}.

\addcontentsline{toc}{section}{Acknowledgments} 
\bigskip 
\par\noindent 
{\textbf{Acknowledgments}} 
\par\noindent 
It is a pleasure to express our thanks to J.L. Lebowitz who suggested the 
problem and to E. Olivieri for many useful discussions and comments. 
We also thank the CMI of Marseille for its kind hospitality and the 
European 
network ``Stochastic Analysis and its Applications" ERB--FMRX--CT96--0075 
for financial support. 
 

\end{document}